\documentclass[12pt,letterpaper]{article}
\pdfoutput=1
\usepackage{jheppub}
\usepackage{amssymb}
\usepackage{amsfonts}
\usepackage{hyperref}
\usepackage{dsfont}
\usepackage{bm}
\usepackage{comment}
\usepackage{mathrsfs}
\usepackage{stmaryrd}

\DeclareMathAlphabet{\mathpzc}{OT1}{pzc}{m}{it}

\newtheorem{tdef}{Definition}

\newcommand{\beq}{\begin{equation}}
\newcommand{\eeq}{\end{equation}}
\newcommand{\bear}{\begin{eqnarray}}
\newcommand{\eear}{\end{eqnarray}}
\newcommand{\set}[1]{\ensuremath{\mathbb{#1}}}
\newcommand*{\defeq}
{\mathrel{\vcenter{\baselineskip0.5ex \lineskiplimit0pt
                     \hbox{\scriptsize.}\hbox{\scriptsize.}}}
                     =}

\newcommand{\proj}[1]{\ensuremath{\mathbb{P}^{#1}}}
\newcommand{\virgolette}{``}

\numberwithin{equation}{section}

\title{One-Dimensional Super Calabi-Yau Manifolds and their Mirrors}

\author[a]{S.~Noja,}
\author[b,c]{S.~L.~Cacciatori,}
\author[b]{F.~Dalla Piazza,}
\author[d,e]{A.~Marrani,}
\author[f]{R.~Re}

\affiliation[a]{Dipartimento di Matematica, Universit\`a degli Studi di Milano,\\
Via Saldini 50, I-20133 Milano, Italy}
\affiliation[b]{Dipartimento di Scienza e Alta Tecnologia,
Universit\`a dell'€™Insubria, \\
Via Valleggio 11, I-22100 Como, Italy.}
\affiliation[c]{INFN, Sezione di Milano,
Via Celoria 16, I-20133 Milano, Italy.}
\affiliation[d]{Centro Studi e Ricerche `Enrico Fermi', Via Panisperna 89A, I-00184 Roma, Italy.}
\affiliation[e]{Dipartimento di Fisica e Astronomia `Galileo Galilei',
Universit\`a di Padova, and INFN, Sezione di Padova, Via Marzolo 8, I-35131 Padova, Italy}
\affiliation[f]{Dipartimento di Matematica e Informatica, Universit\`a degli Studi di Catania,\\
Viale Andrea Doria 6, 95125 Catania, Italy}

\emailAdd{simone.noja@unimi.it}
\emailAdd{sergio.cacciatori@uninsubria.it}
\emailAdd{f.dallapiazza@gmail.com}
\emailAdd{Alessio.Marrani@pd.infn.it}
\emailAdd{riccardo@dmi.unict.it}
\preprint{DFPD/2016/TH-13}

\abstract{We apply a definition of generalised super Calabi-Yau variety (SCY) to supermanifolds of complex dimension one. One of our results is that there are two SCY's having reduced manifold equal to $\mathbb{P}^1$, namely the projective super space $\mathbb{P}^{1|2} $ and the weighted projective super space $\mathbb{WP}^{1|1}_{(2)}$. Then we compute the corresponding sheaf cohomology of superforms, showing that the cohomology with picture
number one is infinite dimensional, while the de Rham cohomology, which is what matters from a physical point of view, remains finite dimensional. Moreover, we provide the complete real and holomorphic de Rham cohomology for generic projective super spaces $\mathbb P^{n|m}$.
We also determine the automorphism groups: these always match the dimension of the projective super group with the only exception of $\mathbb{P}^{1|2} $, whose automorphism group turns out to be larger than the projective general linear supergroup. By considering the cohomology of the super tangent sheaf, we compute the deformations of $\mathbb{P}^{1|m}$, discovering that the presence of a fermionic structure allows for deformations even if the reduced manifold is rigid. Finally,
we show that $\mathbb{P}^{1|2} $ is self-mirror, whereas $\mathbb{WP} ^{1|1}_{(2)}$ has a zero dimensional mirror. Also, the mirror map for $\mathbb{P}^{1|2}$ naturally endows it with a structure of $N=2$ super Riemann surface.}

\keywords{Supergeometry, String Theory, Mirror Symmetry, Calabi-Yau.}

\begin{document}
\maketitle
\flushbottom


\section{Introduction}

\textquotedblleft Super-mathematics\textquotedblright\ has quite a long history, starting from the pioneering papers by Martin, \cite{Martin-1, Martin-2} and Berezin, \cite{Berezin-61, Berezin-66}, before the discovery
of supersymmetry in physics\footnote{Even though anticommutation was proposed yet previously by Schwinger and other physicists, see \cite{storia di Berezin} for a more detailed account.}. After its appearance in physics in the 70s, however,
supergeometry has caught more attention in the mathematical community, and corresponding developments appeared not only in numerous articles but also in devoted books, see \textit{e.g.} \cite{Leites}-\nocite{Alice-Rogers, Bart-Bru-Her,
de Witt, Tuynman, Deligne, Vara}\cite{Fioresi}. In most of the concrete applications of supersymmetry, like in quantum field theory or in supergravity, algebraic properties play a key role, whereas geometry has
almost always a marginal role (apart from the geometric formulation of superspace techniques; see further below). This is perhaps the reason why some subtle questions in supergeometry (see for example \cite{Manin})
have not attracted too much the attention of physicists and, as a consequence, the necessity of further developments has not been stimulated.\medskip

String theory makes exception. Perturbative super string theory is expected to be described in terms of the moduli space of super Riemann surfaces, which results to be itself a supermanifold. However, some ambiguities in
defining super string amplitudes at genus higher than one suggested, already in the 80s, that the geometry of such super moduli space may not be trivially obtained from the geometry of the bosonic underlying space \cite{Atick}.
More than twenty years of efforts have been necessary in order to be able to unambiguously compute genus two amplitudes; \textit{cfr. e.g.} the papers by D'Hoker and Phong \cite{D'Hoker-1}-\nocite{D'Hoker-2, D'Hoker-3, D'Hoker-4,
D'Hoker-5, D'Hoker-6, D'Hoker-7, D'Hoker-8, D'Hoker-9, D'Hoker-10} \cite{D'Hoker-11}, which also include attempts in defining genus three amplitudes, without success but renewing the interest of the physical
community in looking for a solution to the problem of constructing higher genus amplitudes. Through the years, various proposals have been put forward, see \textit{e.g.} \cite{Cacciatori-1}-\nocite{Cacciatori-2,
Cacciatori-3, Piazza-1, Piazza-2, Grushevsky-1, Grushevsky-2, Grushevsky-3, SalvatiManni, Morozov-1, Morozov-2, Matone, Fuchs, Gaberdiel, DuninBarkowski, Poor, Matone-1, DuninBarkowski-1, Matone-2, Matone-3}\cite{Danilov}.

However, most of such constructions were based on the assumption that the supermoduli space is projected (see below for an explanation), but a careful analysis of perturbative string theory and of the corresponding role of
supergeometry \cite{WittenNotes}-\nocite{Witten-1, Witten-2, Witten-3}\cite{Witten-4} suggested that this could not be the case. Indeed, it has been proved in \cite{DonagiWitten} (see also \cite{DonagiWitten-1}) that the
supermoduli space is not split and not projected at least for genus $g\geq 5$. Obviously, such result gave rise to new interest in understanding the peculiarities of supergeometry with respect to the usual geometry, in
particular from the viewpoint of algebraic geometry.\medskip

A second framework in which supergeometry plays a prominent role is the geometric approach to the superspace formalism,
centred on \textit{integral forms} (discussed \textit{e.g.} in \cite{Catenacci, Catenacci-2, CCG-1}; see below), whose application in physics can be traced back to \cite{WittenNotes, Berko}. Superspace techniques are
well understood and used in quantum field theory, supergravity as well as in string theory (see \textit{e.g.} \cite{BW, SS}). They provide a very powerful method to deal with supersymmetric multiplets and to determine
supersymmetric quantities, such as actions, currents, operators, vertex operators, correlators, and so on. However, even when the superspace formulation exists, it is often difficult to extract the component action.
This occurs often in supergravity, in which the superdeterminant of the \textit{supervielbein} is needed for the construction of the action, making the computation pretty cumbersome in a number of cases. On the other hand,
the so-called \textquotedblleft Ectoplasmic Integration Theorem" (EIT) \cite{EIT-1}-\nocite{EIT-2, EIT-3}\cite{EIT-4} can be used in order the extract the component action from the superspace formulation.

Generally, supermanifolds are endowed with a tangent bundle (generated by commuting and anticommuting vector fields) and with an exterior bundle; thus, one would na\"{\i}vely expect the geometric theory of integration on
manifolds to be exported \textit{tout court} in supersymmetric context. Unfortunately, such an extension is not straightforward at all, because top superforms do not exist, due to the fact that the wedge products of the differentials $d\theta$ ($\theta $ being the anticommuting coordinates)
are commuting, and therefore there is no upper bound on the length of the usual exterior $d$-complex. In order to solve this problem, \textit{distribution-like} quantities $\delta (d\theta )$ are introduced, for which a complete Cartan
calculus can be developed. Such distributions $\delta (d\theta )$ then enter the very definition of the \textit{integral forms} \cite{Voronov-1}-\nocite{Voronov-2, Voronov-3, Voronov-4, Belo-1}\cite{Belo-2}, which are a new type
of differential forms requiring the enlargement of the conventional space spanned by the fundamental $1$-forms, admitting distribution-like expressions (essentially, Dirac delta functions and Heaviside step
functions). Within such an extension of the $d$ differential, a complex with an upper bound arises, and this latter can be used to define a meaningful geometric integration theory for forms on supermanifolds. In recent years,
this led to the development of a complete formalism (integral-, pseudo- and super- forms, their complexes and related integration theory) in a number of
papers by Castellani, Catenacci and Grassi \cite{CCG-1, CCG-2, CCG-3}.

In \cite{CCG-2}, the exploitation of integral forms naturally yielded the definition of the Hodge dual operator $\bigstar $ for supermanifolds, by means of the Grassmannian Fourier transform of superforms, which in turn
gave rise to new supersymmetric actions with higher derivative terms (these latter being required by the invertibility of the Hodge operator itself). Such a definition of $\bigstar $ was then converted into a Fourier-Berezin
integral representation in \cite{CCG-4}, exploiting the Berezin convolution. It should also be recalled that integral forms were instrumental in the recent derivation of the superspace action of $D=3$, $N=1 $ supergravity as an integral on a
supermanifold \cite{CCG-6}.

Furthermore, in \cite{CCG-3}, the cohomology of superforms and integral forms was discussed, within a new perspective based on the Hodge dual operator introduced in \cite{CCG-2}. Therein, it was also shown how the
superspace constraints (\textit{i.e.}, the rheonomic parametrisation) are translated from the space of superforms $\Omega ^{(p|0)}$ to the space of integral forms $\Omega ^{(p|m)}$ where $0\leqslant p\leqslant n$, with $n$
and $m$ respectively denoting the bosonic and fermionic dimensions of the supermanifold; this naturally let to the introduction of the so-called Lowering and Picture Raising Operators (namely, the Picture Changing
Operators, acting on the space of superforms and on the space of integral forms), and to their relation with the cohomology.

In light of these achievements, integral forms are crucial in a consistent geometric (superspace) approach to supergravity actions. It is here worth remarking that in \cite{CCG-1} the use of integral forms, in the framework
of the group manifold geometrical approach \cite{gm-1, gm-2} (intermediate between the superfield and the component approaches) to supergravity, led to the proof of the aforementioned EIT, showing that the origin of that formula
can be understood by interpreting the superfield action itself as an integral form. Subsequent further developments dealt with the construction of the super Hodge dual, the integral representation of Picture Changing
Operators of string theories, as well as the construction of the super-Liouville form of a symplectic supermanifold \cite{CCG-5}.\medskip

A third context in which super geometry may be relevant is mirror symmetry. In \cite{Sethi}, Sethi proposed that the extension of the concept of mirror symmetry to super Calabi-Yau manifolds (SCY's) could improve the definition
of the mirror map itself, since supermanifolds may provide the correct mirrors of rigid manifolds. Such a conjecture has been strengthened by the works of Schwarz \cite{Schwarz-1995, Schwarz-1995-1} in the early days, but
it seems to have been almost ignored afterwards, at least until the paper of Aganagic and Vafa \cite{AgaVafa} in 2004, in which a general super mirror map has been introduced and, in particular, it has been shown that the
mirror of the super Calabi-Yau space $\mathbb{P}^{3|4}$ is, in a suitable limit, a quadric in $\mathbb{P}^{3|3}\times \mathbb{P}^{3|3}$. This is a quite interesting case, since these SCY's are related to amplitude
computations in (super) quantum field theories, see \textit{e.g.} \cite{WittenTwistor}. Since then, a number of studies on mirror symmetry for SCY's has been carried on, see for example \cite{Rocek-2004}-\nocite{Ahn-2005, Ahl-2006}\cite{Garavuso-2011}.
However, a precise definition of SCY is currently missing, and, consequently, the definition of mirror symmetry and its consequences is merely based on physical intuition.

\

The aim of the present paper is to provide a starting point for a systematic study of SCY's, by addressing the lowest dimensional case: SCY's whose bosonic reduction has complex dimension one.

\noindent In section \ref{sec:superprojective} we collect some definitions in supergeometry and introduce the projective super spaces, which will play a major role in what follows. We will not dwell into a detailed exposition,
and we address the interested reader \textit{e.g.} to \cite{Deligne} and \cite{Manin} for a mathematically thorough treatment of supergeometry. We also recall that an operative exposition of supergeometry, aimed at
stressing its main connections with physics, is given in \cite{WittenNotes}.

\noindent In section \ref{geometry and coomology} we will be concerned with the geometry of the projective super space $\mathbb{P}^{1|2}$ and of the weighted projective super space $\mathbb{WP}_{(2)}^{1|1}$. \v{C}ech and de Rham cohomology
of super differential forms are computed for these super varieties: here some interesting phenomena occur. Indeed we will find that on the one hand one there might be some infinite-dimensional \v{C}ech cohomology groups as soon as one deals with more
than one odd coordinate (as in the case of $\mathbb{P}^{1|2}$); on the other hand this pathology gets cured at the level of de Rham cohomology, where no infinite dimensional groups occur. Our interest in these two particular supermanifolds originates from the fact that, together with the class of the so-called \emph{N=2 super Riemann surfaces} ($N=2$ SRS's) which will be shortly addressed in what follows, $\mathbb{P}^{1|2}$ and $\mathbb{WP}_{(2)}^{1|1}$ are indeed the \emph{unique} (non-singular) SCY's\footnote{In the sense specified further below.} having reduced manifold given by $\mathbb{P}^{1}$. These are therefore the simplest candidates to be considered, as one is interested into extending the mirror symmetry construction in dimension $1$ to a super geometric context, pursuing a task initially suggested in \cite{Sethi}. Moreover, despite we keep our attention to the case $n|m=1|2$
we also provide the de Rham cohomology of projective super spaces having generic dimension.

\noindent In section \ref{sec:mirror map} we will then construct the mirrors of the projective super spaces $\mathbb{P}^{1|2}$ and $\mathbb{WP}_{(2)}^{1|1}$, following a recipe introduced in \cite{AgaVafa}. Moreover, we will show that, surprisingly, by
means of the mirror construction, $\mathbb{P}^{1|2}$ actually gives a concrete example of $N=2$ SRS. \\
Finally, the main results and perspectives for further developments are discussed in section \ref{sec: conclusions}, whereas an appendix is devoted to
illustrating the coherence of the adopted rule of signs.


\section{Supermanifolds and Projective Super Spaces}

\label{sec:superprojective}


\subsection{Definitions and Notions in Supergeometry}

In general, the mathematical basic notion that lies on the very basis of any physical supersymmetric theory is the one of $\mathbb{Z} _2$-grading: algebraic constructions such as rings, vector spaces, algebras and so on,
have their $\mathbb{Z} _2$-graded analogues, usually called in physics \emph{super rings, super vector spaces} and \emph{super algebras}, respectively. \newline
A ring $(A, +, \cdot)$, for example, is called a super ring if $(A, +)$ has two subgroups $A_0$ and $A_1$, such that $A = A_0 \oplus A_1$ and
\begin{equation}
A_i \cdot A_j \subset A_{(i + j)\mbox{\emph{\scriptsize{mod}}}\,2} \qquad
\forall\, i , j \in \mathbb{Z} _2.
\end{equation}
\noindent The generalisation of vector spaces to super vector spaces, as well as of algebras to super algebras, follows the same lines.\newline
Given an homogeneous element with respect to the $\mathbb{Z} _2$-grading of a super ring, we can define an application, called \emph{parity} of the element, as follows:
\begin{equation}
a \longmapsto |a| \defeq \left \{
\begin{array}{lr}
0 & \qquad a \in A_0 \\
1 & \qquad a \in A_1%
\end{array}
\right. .
\end{equation}
Elements such that $|a| = 0$ ($a \in A_0$) are called \emph{even} or \emph{bosonic}, and elements such that $|a| = 1$ ($a \in A_1$) are called \emph{odd} or \emph{fermionic}.\newline
It should be pointed out that so far no supersymmetric structure has been related to any super commutativity of elements, provided by the \emph{super commutator}, i.e. a bilinear map acting on two generic homogeneous
elements $a, b$ in a super ring $A$, as follows :
\begin{equation}
(a, b) \longmapsto a\cdot b - (-1)^{|a| \cdot |b|} \,b\cdot a.
\end{equation}
By additivity, this extend to a map $[ \cdot, \cdot ] : A \times A
\rightarrow A$.\newline
By definition, a super ring is \emph{super commutative} if and only if \emph{all} the super commutators among elements vanish (or in other words, if and only if the center of the super ring is the super ring itself); thus, on the homogeneous elements it holds that $a \cdot b = (-1)^{|a| \cdot |b|} \, b \cdot a,$ for all $a \in A_i, b \in A_i $ with $i \in \mathbb{Z} _2.$ Supergeometry only deals with this class of super rings, allowing for anti-commutativity of odd elements. This
has the following obvious fundamental consequence: \emph{all odd elements are nilpotent}.\newline
A basic but fundamental example of super commutative ring (actually algebra) is provided by the polynomial superalgebra over a certain field $k$, denoted as $k[x_1, \ldots, x_p , \theta_1, \ldots, \theta_q]$, where $x_1,
\ldots, x_p $ are even generators, and $\theta_1, \ldots, \theta_q$ are odd generators. The presence of the odd anti commuting part implies the following customary picture for this super algebra:
\begin{eqnarray}
k[x_1, \ldots , x_p, \theta_1, \ldots, \theta_q] \cong k[x_1, \ldots, x_p ]
\otimes_k \bigwedge[\theta_1, \ldots, \theta_q]
\end{eqnarray}
which makes apparent that the theta's are generators of a Grassmann algebra. Even and odd superpolynomials might be expanded into the odd (and therefore nilpotent) generators as follows
\begin{eqnarray}
P_{even} (x, \theta) = f_0 (x) + \sum_{i<j =1}^q f_{ij} (x) \theta_i
\theta_j + \sum_{i<j<k<l=1}^q f_{ijkl} (x)\theta_i \theta_j \theta_k
\theta_l +\ldots
\end{eqnarray}
\begin{eqnarray}
P_{odd} (x, \theta) = \sum_{i =1}^q f_i (x) \theta_i + \sum_{i<j<k =1}^q
f_{ijk} (x) \theta_i \theta_j \theta_k + \ldots
\end{eqnarray}
where the $f$'s are usual polynomials in $k[x_1, \ldots, x_p]$ and we have written $\theta_i \theta_j $ instead of $\theta_i \wedge \theta_j$ for the sake of notation.

As one wishes to jump from pure algebra to geometry, in physics it is customary to consider a supermanifold $\mathpzc{M}$ of dimension $p|q$ (that is, of even dimension $p$ and odd dimension $q$) as described locally by $p$
even coordinates and $q$ odd coordinates, as a generalisation of the standard description of manifolds in differential geometry. Even if this is feasible \cite{WittenNotes}, due to the presence of nilpotent elements, in supergeometry it is preferable to adopt an algebraic geometric oriented point of view, in which a supermanifold is conceived as a \emph{locally ringed space} \cite{Deligne} \cite{Manin} \cite{Vara} \cite{DonagiWitten}.

Within this global point of view, we define a \emph{super space} $\mathpzc{M}$ to be a $\mathbb{Z} _2$-graded locally ringed space, that is a pair $(|\mathpzc{M}|, \mathcal{O}_{\mathpzc{M}})$, consisting of a topological space
$|\mathpzc{M}|$ and a sheaf of super algebras $\mathcal{O}_{\mathpzc{M}}$ over $|\mathpzc{M}|$, such that the stalks $\mathcal{O}_{\mathpzc{M}, x}$ at every point $x \in |\mathpzc{M}|$ are local rings. Notice that this makes
sense as a requirement, for the odd elements are nilpotent and this reduces to ask that the even component of the stalk is a usual local commutative ring. \newline
Morphisms between super spaces become morphisms of locally ringed spaces, i.e. they are given by a pair
\begin{eqnarray}
(\phi, \phi^\sharp) : (|\mathpzc{M}|, \mathcal{O}_{\mathpzc{M}})
\longrightarrow (|\mathpzc{N}|, \mathcal{O}_{\mathpzc{N}}),
\end{eqnarray}
where $\phi : |\mathpzc{M}| \rightarrow |\mathpzc{N}|$ is a continuous function and $\phi^\sharp : \mathcal{O}_{\mathpzc{N}} \rightarrow \phi_*\mathcal{O}_{\mathpzc{M}}$ is morphism of sheaves (of super rings or super
algebras), which needs to preserve the $\mathbb{Z} _2$-grading.\newline
Clearly, $\mathcal{O}_{\mathpzc{M}}$ contains the subsheaf $\mathcal{J}_{\mathpzc{M}}$ of ideals of all nilpotents, which is generated by all odd elements of the sheaf: this allows us to recover a \emph{purely even}
super space, $(|\mathpzc{M}|, {\raisebox{.1em}{$\mathcal{O}_{\mathpzc{M}}$}\left/\raisebox{-.1em}{$\mathcal{J}_{\mathpzc{M}}$}\right.})$, which is called \emph{reduced space} $\mathpzc{M}_{red}$ underlying $\mathpzc{M}$.
There always exists a closed embedding $\mathpzc{M}_{red} \hookrightarrow \mathpzc{M}$, given by the morphism $(id_{|\mathpzc{M}|},
i^{\sharp}) : (|\mathpzc{M}|, {\raisebox{.1em}{$\mathcal{O}_{\mathpzc{M}}$} \left/\raisebox{-.1em}{$\mathcal{J}_{\mathpzc{M}}$}\right.}) \rightarrow (|\mathpzc{M}|, \mathcal{O}_{\mathpzc{M}})$, where $i^{\sharp} : \mathcal{O}_{\mathpzc{M}}
\rightarrow id_{|\mathpzc{M}|*}{\raisebox{.1em}{$\mathcal{O}_{\mathpzc{M}}$}\left/\raisebox{-.1em}{$\mathcal{J}_{\mathpzc{M}}$}\right.} = {\raisebox{.1em}{$\mathcal{O}_{\mathpzc{M}}$}\left/\raisebox{-.1em}{$\mathcal{J}_{\mathpzc{M}}$}\right.}$.

A special super space can be constructed as follows: given a topological space $|\mathpzc{M}|$ and a locally free sheaf of $\mathcal{O}_{|\mathpzc{M}|}$-modules $\mathcal{E}$, one can take $\mathcal{O}_{\mathpzc{M}}$ to be the
sheaf $\bigwedge^{\bullet} \mathcal{E}^\vee$: this makes $\mathcal{O}_{\mathpzc{M}}$ out of a super commutative sheaf whose stalks are local rings. Similarly to \cite{DonagiWitten}, super spaces constructed in this
way are denoted as $\mathfrak{S}(|\mathpzc{M}|, \mathcal{E})$.\newline
Examples of this construction are \emph{affine super spaces} $\mathbb{A}^{p|q} \defeq \mathfrak{S} (\mathbb{A} ^p, \mathcal{O}_{\mathbb{A}^p}^{\oplus q})$: in this case, $\mathbb{A} ^p$ is the ordinary $p$-dimensional
affine space over $\mathbb{A} $ and $\mathcal{O}_{\mathbb{A} ^p}$ is the trivial bundle over it. Super spaces like these are common in supersymmetric field theories, where one usually works with $\mathbb{R} ^{p|q}$ or $\mathbb{C} ^{p|q}.$

A \emph{supermanifold} is defined as a super space which is \emph{locally} isomorphic\footnote{In the $\mathbb{Z} _2$-graded sense: here indeed isomorphisms are isomorphisms of super algebras.} to $\mathfrak{S} (|\mathpzc{M}|, \mathcal{E})$
for some topological space $|\mathpzc{M}|$ and some locally free sheaf of $\mathcal{O}_{|\mathpzc{M}|}$-module $\mathcal{E}$. \newline
Along this line of reasoning, then, one recovers (out of a globally
defined object!) the original differential geometric induced view that physics employs, in which a \emph{real} supermanifold of dimension $p|q$ locally resembles $\mathbb{R} ^{p|q}$ and, analogously, a \emph{complex} supermanifold of
dimension $p|q$ locally resembles $\mathbb{C} ^{p|q}$, defined above: the gluing data are encoded in the cocycle condition satisfied by the structure sheaf. \newline
Given a supermanifold $\mathpzc{M}$, we will call $\mathpzc{M}_{red}$ the pair $(|M|, {\raisebox{.1em}{$\mathcal{O}_{\mathpzc{M}}$}\left/\raisebox{-.1em}{$\mathcal{J}_{\mathpzc{M}}$}\right.})$, which is an
ordinary manifold conceived as a locally ringed space of a certain type: as above, a closed embedding $\mathpzc{M}_{red} \hookrightarrow \mathpzc{M}$ will always exist. \newline
It is here worth pointing out that, on the contrary, the definition of supermanifold does not imply the existence of a projection $\mathpzc{M} \rightarrow \mathpzc{M}_{red}$: this would correspond to a morphism of $(id_{|\mathpzc{M}|},
\pi^{\sharp}) : (|\mathpzc{M}|, \mathcal{O}_{\mathpzc{M}}) \rightarrow (|\mathpzc{M}| , {\raisebox{.1em}{$\mathcal{O}_{\mathpzc{M}}$}\left/\raisebox{-.1em}{$\mathcal{J}_{\mathpzc{M}}$}\right.})$, where $\pi^\sharp $
is a sheaf morphism that embeds ${\raisebox{.1em}{$\mathcal{O}_{\mathpzc{M}}$}\left/\raisebox{-.1em}{$\mathcal{J}_{\mathpzc{M}}$}\right.}$ into $\mathcal{O}_{\mathpzc{M}}$; moreover, one should also endow the sheaf $\mathcal{O}_{\mathpzc{M}}$ with
the structure of sheaf of $\mathcal{O}_{\mathpzc{M}_{red}} $-modules. When such a projection exists, the supermanifold
is said to be \emph{projected}. Conceiving a supermanifold in terms of the gluing data between open sets covering the underlying topological space, the projectedness of such a supermanifold implies the even transition
functions to be written as functions of the ordinary local coordinates on the reduced manifold only: there are no nilpotents (e.g. bosonisation of odd elements) at all. Obstruction to the existence of such a projection for the
case of the supermoduli space of super Riemann surfaces has been studied in \cite{DonagiWitten} and it is an issue that has striking consequences in superstring perturbation theory, as mentioned in the introduction. \newline
A stronger condition is realised when the supermanifold is \emph{globally} isomorphic to some local model $\mathfrak{S}(|\mathpzc{M}|, \mathcal{E})$. Such supermanifolds are said to be \emph{split}. If this is the case, not
only the even transition functions have no nilpotents, but the odd transition functions can be chosen in such a way that they are linear in the odd coordinates. This bears a nice geometric view of split supermanifolds:
they can be globally regarded as a vector bundle $\tilde{\mathcal{E}} \rightarrow \mathpzc{M}_{red}$ on the reduced manifold having purely odd fibers, as the above definition of supermanifold suggests by itself.


\subsection{Projective Super Spaces and Weighted Projective Super Spaces}

The supermanifolds known as (complex) projective super spaces, denoted by $\mathbb{P} ^{n|m}$, have been discussed extensively in the literature and introduced from several different points of view, both in mathematics and in
physics, being of fundamental importance in twistor string theory.

In most cases, complex projective spaces are regarded as a quotient of the super spaces $\mathbb{C} ^{n|m}$ by the even multiplicative group $\mathbb{C}^\times$, so realising a super analogue of the set of homogeneous coordinates
$[X_1 : \ldots : X_n : \Theta_1 : \ldots : \Theta_m]$ obeying $[X_1 : \ldots : X_n : \Theta_1 : \ldots : \Theta_m] = [\lambda X_1 : \ldots : \lambda X_n : \lambda \Theta_1 : \ldots : \lambda \Theta_m]$, where $\lambda \in \mathbb{C}^\times$
(see for example \cite{DonagiWitten}, \cite{WittenNotes}). In contrast with this global construction by a quotient, a popular local construction realises $\mathbb{P} ^{n|m}$ mimicking the analogous constructions of $\mathbb{P} ^n$ as a complex manifold, namely by specifying it as $n+1$ copies of $\mathbb{C} ^{n|m}$ glued together by the usual relations. This construction relies on the possibility to pair the usual bosonic local coordinates with $q$ fermionic anticommuting local
coordinates: such an intuitive approach can be made rigorous using the functor of points formalism \cite{Fioresi}. A more rigorous treatment, connecting the requested invariance under the action on $\mathbb{C} ^\times$
with the structure of the sheaf of super commutative algebra characterising the projective super space, can be found in \cite{Catenacci}.

An elegant construction of the projective super space consistent with the notions introduced above is given in \cite{Manin}. $\mathbb{P}^{n|m}$ can actually be presented as a (split) complex supermanifold, as
follows. We consider a super $\mathbb{C} $-vector space $V = V_0 \oplus V_1$ of rank $n+1|m$. As one can imagine, the topological space underlying the super projective space coincides with the usual one and it is given by the
projectivization of the even part of $V$, called simply $\mathbb{P} ^n $. This tells that $\mathbb{P} ^n$ can be covered by $n+1$ open sets $\{U_i\}_{i = 0, \ldots, n} $, characterised by
\begin{eqnarray}
U_i \defeq \left \{ [X_0 : \ldots : X_n ] \in \mathbb{P} ^n : X_i \neq 0
\right \},
\end{eqnarray}
in such a way that one can form a system of local affine coordinates on $U_i$ given by $z^{(i)}_j \defeq {\raisebox{.1em}{$X_j$}\left/\raisebox{-.1em}{$X_i$}\right.}$ for $j\neq i$. Intuitively, as above, we would like to have something
similar for the odd part of the geometry: this is achieved by realising a sheaf of super algebras on $\mathbb{P} ^n$, as follows:
\begin{eqnarray}
U_i \longmapsto \left (\bigoplus_{\ell = 0}^m \bigwedge^\ell V_1^\vee
\otimes \mathcal{O}_{\mathbb{P} ^n} (-1) \right ) (U_i).
\end{eqnarray}
This is isomorphically mapped to the structure sheaf $\mathcal{O}_{\mathbb{P}^{n|m}}$ of the projective super space, by a map induced by
\begin{eqnarray}
V_1^\vee \otimes \mathcal{O}_{\mathbb{P} ^n} (-1) (U_i) \owns \Theta_\alpha \otimes X_i^{-1} \longmapsto \theta^{(i)}_\alpha \defeq \frac{\Theta_\alpha}{X_i} \in \mathcal{O}_{\mathbb{P} ^{n|m}} (U_i) ,
\end{eqnarray}
where it should be stressed that $\Theta_\alpha $ is a generator for $V_1^\vee$, $X^{-1}_i $ is a section of $\mathcal{O}_{\mathbb{P} ^n} (-1)$ over $U_i$ and the $\theta^{(i)}_\alpha$, where $\alpha = 1, \ldots, m$ are promoted to
local odd coordinates over $U_i$ for the projective super space $\mathbb{P}^{n|m}$.\newline
This construction makes apparent that, in the notation introduced in the previous section, $\mathbb{P} ^{n|m} = \mathfrak{S} (\mathbb{P} ^n, V_1 \otimes \mathcal{O}_{\mathbb{P} ^n} (1))$. \newline
One can also read out the transition rules on $U_i \cap U_j$, even and odd, that are usually written as:
\begin{align}
z^{(i)}_k = \frac{z^{(j)}_k}{z^{(j)}_i}, \qquad \qquad \theta^{(i)}_\alpha = \frac{\theta^{(j)}_\alpha}{z^{(j)}_i}.
\end{align}
In the language of morphisms of ringed spaces, we would have an isomorphism
\begin{align}
(\phi_{U_i \cap U_j}, \phi^\sharp_{U_i \cap U_j}) : (U_i \cap U_j, \mathcal{O}_{\mathbb{P} ^{n|m}} (U_i )\lfloor_{U_j} ) \longrightarrow (U_i \cap U_j, \mathcal{O}_{\mathbb{P} ^{n|m}} (U_j )\lfloor_{U_i}),
\end{align}
with $\phi_{U_i \cap U_j} : U_i \cap U_j \rightarrow U_i \cap U_j$ being the usual change of coordinate on projective space and $\phi^\sharp_{U_i \cap U_j} : \mathcal{O}_{\mathbb{P} ^{n|m}} (U_j \lfloor_{U_i}) \rightarrow
\left
( (\phi_{{U_i \cap U_j}})_{*} \mathcal{O}_{\mathbb{P} ^{n|m}} \right) (U_i \lfloor_{U_j}) $, so that
\begin{eqnarray}
\left (\phi_{U_i \cap U_j} \right )_{*} (z^{(i)}_k) = \frac{z^{(j)}_k}{z^{(j)}_i}, \qquad \qquad \left (\phi_{U_i \cap U_j} \right )_{*} (\theta^{(i)}_\alpha ) = \frac{\theta^{(j)}_\alpha}{z^{(j)}_i}.
\end{eqnarray}
We note, incidentally, that the cocycle relation is indeed satisfied.\vspace{4pt}\newline

Before proceeding further with our treatment, we here generalise a little the construction above, allowing us to deal in a somehow unified way with the weighted projective super spaces, as well. We will actually be interested in the case in which the odd part of the
geometry carries different weights compared to the even part, which is made by an ordinary projective space. \newline
Since above we considered $V= V_0 \oplus V_1$ to be a super $\mathbb{C} $-vector space, then it yields a well defined notion of dimension, namely $n+1|m$, and one can actually take a basis for it. Focusing on the odd part, we take $\{\Theta_\alpha \}_{\alpha =1, \ldots, m }$ as a system of generators for $V_1$. Then, we might realise a more general sheaf of super algebras by
\begin{eqnarray}
U_i \longmapsto \bigwedge^\bullet \left ( \bigoplus_{\alpha=1}^m\Theta_\alpha^\vee \otimes \mathcal{O}_{\mathbb{P} ^n} (-{w}_\alpha) \right ) (U_i).
\end{eqnarray}
In other words, each odd variable has been assigned a weight $w_\alpha$, which affects the transition functions: the ordinary case of $\mathbb{P} ^{n|m}$ is recovered assigning $w_\alpha = 1$ for each $\alpha = 1, \ldots, m$.\newline
We will call this space weighted projective super space and we will denote it by $\mathbb{WP} ^{n|m}_{(w_1, \ldots , w_m)}$, where the string $(w_1, \ldots, w_m)$ gives the fermionic weights. In this paper, we will be
particularly concerned with low dimensional examples of projective and weighted projective super space, namely $\mathbb{P} ^{1|2}$ and $\mathbb{WP}^{1|1}_{(2)}$, whose geometry will be studied in some details in the following section.


\subsection{Vector Bundles over $ \pmb{\mathbb{P} ^1}$, Grothendieck's Theorem and cohomology of $\pmb {\mathcal{O}_{\mathbb{P} ^n}(k)}$-bundles}

In this section we recall and comment a classification result due to Grothendieck, which will then be exploited to study projective super spaces whose reduced space is $\mathbb{P} ^1.$ Moreover, for future
use, the cohomology of $\mathcal{O}_{\mathbb{P} ^n}(k)$-bundles is given. \vspace{4pt}\newline

The main result about vector bundles on $\mathbb{P} ^1$ is that any holomorphic vector bundle $\mathcal{E}$ of rank $n$ is isomorphic to the direct sum of $n$ line bundles and the decomposition is unique up to permutations of such line
bundles, that is :
\begin{eqnarray}
\mathcal{E} \cong \bigoplus_{i = 1}^n \mathcal{O}_{\mathbb{P}^{1 }} (k_i) ,
\end{eqnarray}
where the ordered sequence $k_1 \geq k_2 \geq \ldots \geq k_n$ is uniquely determined (see \cite{GH} for a complete proof). We will refer at it as Grothendieck's Theorem. Basically, it states that the only interesting
vector bundles on $\mathbb{P}^{1 }$ are the line bundles on it, which in turn are all of the form $\mathcal{O}_{\mathbb{P}^{1 }} (k) $ for some $k \in \mathbb{Z}$ (recall that $\mbox{Pic} (\mathbb{P}^{1} ) \cong \mathbb{Z} $).\newline
Concretely, since every (algebraic) vector bundle over $\mathbb{C} $ is trivial, the restriction of a vector bundle $\mathcal{E}$ over $\mathbb{P}^{1} $ of rank $n$ to the standard open sets $U_0 \defeq \{[X_0 : X_1] : X_0
\neq 0 \} \cong \mathbb{C} $ and $U_1 \defeq \{ [X_0:X_1] : X_1 \neq 0\} \cong \mathbb{C} $ is trivial. Choosing the coordinates $z \defeq \frac{X_1}{X_0}$ on $U_0$ and $w \defeq \frac{X_0}{X_1} = \frac{1}{z}$ on $U_1$, an isomorphism $\mathcal{E}\lfloor_{U_0} \rightarrow \mathcal{O}_{\mathbb{P}^{1} }^{\oplus n} \lfloor_{U_0}$ is equivalent to an isomorphism of $\mathcal{O}_{\mathbb{P}^{1} }(U_0) = \mathbb{C} [z]$-modules, as follows :
\begin{eqnarray}
\phi_0 : \mathcal{E}(U_0) \overset{\cong}{\longrightarrow} \mathcal{O}_{\mathbb{P}^{1} } (U_0)^{\oplus n} = \mathbb{C} [z]^{\oplus n}.
\end{eqnarray}
Analogously, the following isomorphism of $\mathcal{O}_{\mathbb{P}^{1} }(U_1)=\mathbb{C} \left [ z^{-1} \right ]$-modules holds :
\begin{eqnarray}
\phi_1 : \mathcal{E}(U_1) \overset{\cong}{\longrightarrow} \mathcal{O}_{\mathbb{P}^{1} } (U_1)^{\oplus n} = \mathbb{C} \left [z^{-1} \right]^{\oplus n}.
\end{eqnarray}
Clearly, two such isomorphisms $\phi_i $ and $\phi_i^\prime$ yields to an automorphism $\phi_i \circ \phi_i^\prime : \mathbb{C} [t]^{\oplus n} \rightarrow \mathbb{C} [t]^{\oplus n}$ where $t = z $ for $i = 0$ and $t=z^{-1}$ for $i= 1,$
so $\phi_i \circ \phi_i^\prime$ determines an invertible $n \times n$ matrix, having coefficients in $\mathbb{C} [t]$.\newline
The composition $\phi_{01} \defeq \phi_0 \circ \phi_{1}^{-1}$ gives the glueing relation between the two trivial bundles over $U_0$ and $U_1$: it is again given by an invertible $n \times n$ matrix having coefficient in
$\mathbb{C} [z, z^{-1}]$, with determinant $z^{k}$ for some $k\in \mathbb{Z} $, up to a non-zero constant. Thus, classifying rank $n$ vector bundles over $\mathbb{P}^{1 } $ corresponds to classifying invertible
matrices $M \in GL (n, \mathbb{C} [z, z^{-1}])$ up to the following equivalence:
\begin{eqnarray}
M(z, z^{-1}) \, \sim \, A (z) M (z, z^{-1}) B (z^{-1}) \qquad A (z) \in GL(n, \mathbb{C} [z]) , B (z^{-1}) \in GL (n, \mathbb{C} [z^{-1}]).  \notag
\end{eqnarray}
By a theorem due to Birkhoff, $M(z, z^{-1})$ belongs to the same class of a \emph{diagonal} matrix $M_{d} = \mbox{diag} (z^{k_1}, \ldots, z^{k_n})$ where $k_i \in \mathbb{Z} .$ Therefore any bundle over $\mathbb{P}^{1} $ is isomorphic
to a direct sum of line bundles $\mathcal{O}_{\mathbb{P}^{1 }} (k_1) \oplus \ldots \oplus \mathcal{O}_{\mathbb{P}^{1}} (k_n) $.\vspace{4pt}\newline

By looking at vector bundles over $\mathbb{P}^{1} $ as sheaves of locally free $\mathcal{O}_{\mathbb{P}^{1} }$-modules, the theorem reduces the problem of computing sheaf cohomology over $\mathbb{P}^{1} $ to computing
the sheaf cohomology of $\mathcal{O}_{\mathbb{P}^{1 }} (k) $, which is well-known. In general, for $k \geq 0$ one has $H^0(\mathbb{P}^n, \mathcal{O}_{\mathbb{P}^n} (k)) = \mathbb{C}[x_0, \ldots, x_n]^{(k)}$, the \emph{degree-k linear
subspace of the polynomial ring}, therefore
\begin{eqnarray}
h^0 (\mathbb{P}^n, \mathcal{O}_{\mathbb{P}^n} (k)) = \left (\begin{array}{c}
k+n \\
k
\end{array}
\right ) = \frac{(n+k)!}{k! \cdot n!} \qquad k \geq 0,
\end{eqnarray}
and if $k < 0$ one has $H^n( \mathbb{P}^n, \mathcal{O}_{\mathbb{P}^n} (k)) = \big \langle x_0^{i_0} \cdot \ldots \cdot x^{i_n}_n : i_j < 0, \quad \sum_{i=0}^{n} i_j = k \big \rangle_{\mathbb{C}}$. It is an exercise in
combinatorics to see that
\begin{eqnarray}
h^n (\mathbb{P}^{n} , \mathcal{O}_{\mathbb{P}^{n} } (k)) = {\binom{{|k| -1}}{{|k| - n -1} }} \qquad k <0, \; |k| \geq n + 1.
\end{eqnarray}
These results will be used to compute the cohomology in the following section.


\section{The Geometry and Cohomology of $\pmb{\mathbb{P}^{1|2}}$ and $\pmb{\mathbb{WP}^{1|1}_{(2)}}$}

\label{geometry and coomology}


\subsection{Super Calabi-Yau Varieties}

The physical approach to Calabi-Yau geometries in a supersymmetric context is based on the differential geometric point of view, defining a super Calabi-Yau manifold (SCY) as a Ricci-flat
supermanifold. Indeed, a generalisation of tensor calculus to a supersymmetric context exists, making use of the notion of super Riemannian manifold, super curvature tensor and super Ricci tensor. \newline
The crucial observation concerning projective super spaces is that there exists a generalisation of the Fubini-Study metric to the supersymmetric context. Considering $\mathbb{P}^{n|m} $, one can define the super
K\"ahler potential
\begin{eqnarray}
\mathcal{K}^s = \log \left (\sum_{i= 0 }^n X_i \bar X_i + \sum_{j =1}^m \Theta_j \bar \Theta_j \right )
\end{eqnarray}
everywhere on $\mathbb{P}^{n|m} $, which reduces to the ordinary Fubini-Study potential as one restrict it to the underlying reduced manifolds. Locally, on a patch of the projective super space, it takes the form
\begin{eqnarray}
K^s = \log \left (\sum_{i= 1}^n z_i \bar z_i + \sum_{j =1}^m \theta_j \bar \theta_j \right ).
\end{eqnarray}
Notice that, in this complex differential geometric context, the variables are paired with their anti-holomorphic counterparts, as customary in theoretical physics. \newline
The super K\"ahler form is defined in the local patch to be
\begin{eqnarray}
\Omega^s \defeq \partial \bar \partial K^s \quad \mbox{or analogously} \quad \Omega^s \defeq \partial_A \partial_{\bar B} K^s dX^A dX^{\bar B}
\end{eqnarray}
where $\partial $ and $\bar \partial$ are the holomorphic and anti-holomorphic super derivatives; we refer to the Appendix A for details. Then, the super metric tensor is simply given by
\begin{eqnarray}
H^s_{A \bar B} \defeq \partial_A \partial_{\bar B} K^s,
\end{eqnarray}
and the Ricci tensor reads
\begin{eqnarray}
\mbox{Ric}_{A \bar B} \defeq \partial_A \partial_{\bar B} \log \left ( \mbox{Ber} H^s \right ).
\end{eqnarray}
Notice that, in comparison to the ordinary complex geometric case, the unique modification is that the determinant of the metric has been substituted by the Berezinian \cite{Deligne} \cite{WittenNotes} of the super metric.\newline
Remarkably, as one chooses the projective super spaces of the form $\mathbb{P}^{n|n+1} $ for $n\geq1$ (note that $\mathbb{P}^{0|1} \cong \mathbb{C}^{0|1} $), endowed with the super Fubini-Study metric defined as above, then
one gets a vanishing super Ricci tensor!\newline
We stress that, as it is common in the context of super geometry, even an easy calculation might present some difficulties, due to the anti-commutativity of some variables. One needs to establish and keep
coherent conventions throughout the calculations. As an example, a detailed computation of the vanishing of the super Ricci tensor in the case of $\mathbb{P}^{1|2} $ is reported in Appendix A. The same calculation can be
easily generalised to any $\mathbb{P}^{n|n+1} $ for $n\geq 2$.\vspace{4pt}

Actually, defining a SCY manifold by requiring that its super Ricci tensor vanishes turns out to be a very strong request. Moreover, it is not that useful, for it is often hard to write down super metrics for interesting classes of supermanifolds: for example, there is not straightforward generalisation of the super Fubini-Study metric to the case of weighted projective super spaces. Moreover, by the result in \cite{Rocek-2004} a Ricci-flat K\"{a}hler supermetric on $\mathbb{WP}_{(2)}^{1|1}$ does not exist. Still, it is possible to give a weaker, but certainly more useful, definition:

\begin{tdef}
We say that an orientable super projective variety $X$ is super Calabi-Yau if its Berezinian sheaf is trivial, that is $\mathcal{B}\mathpzc{er}_X \cong \mathcal{O}_X$.
\end{tdef}

It is here worth remarking that this definition is again the super analogue of the usual algebraic geometric definition of an ordinary CY variety, that calls for a trivial canonical sheaf. Indeed, the Berezinian sheaf is, in some sense (see \cite{Deligne} \cite{Manin} \cite{WittenNotes}),
the super analogue of the canonical sheaf, since the sections of the Berezinian transform as densities and they are the right objects to define a meaningful notion of super integration, i.e. the Berezin integral. In other words, then, a
SCY variety is one whose Berezinian sheaf has an everywhere non-zero global section.

We now start using of Grothendieck's Theorem in order to prove the triviality of the Berezinian sheaf of the supermanifolds $\mathbb{P}^{1|2} $ and $\mathbb{WP} ^{1|1}_{(2)}$, and hence to confirm that they are both SCY
varieties. To compute the cohomologies, we will consider the varieties as split supermanifolds; in this case, this amounts to consider the total space of vector bundles over $\mathbb{P}^{1} $ (the
reduced manifold, having odd fibers). Then, we will achieve the splitting into line bundles over $\mathbb{P}^{1} $ and we will compute their cohomology.\newline
We start considering $\mathbb{P}^{1|2} $. Following the above line of reasoning, we have two patches, namely $U_z $ and $U_w$: switching from one patch to the other yields the following transformations
\begin{align}
\left \{
\begin{array}{lcl}
w \longmapsto z = \frac{1}{w} &  &  \\
\phi_0 \longmapsto \theta_0 = \frac{\phi_1}{w} &  &  \\
\phi_1 \longmapsto \theta_1 = \frac{\phi_2}{w} &  &
\end{array}
\right. .
\end{align}
The structure sheaf, $\mathcal{O}_{\mathbb{P}^{1|2} }$, is therefore locally generated by
\begin{eqnarray}
\mathcal{O}_{\mathbb{P}^{1|2} } (U_z) = \big \langle 1 \big \rangle_{\mathcal{O}_{\mathbb{P}^{1|2} }(U_z)} = \big \langle 1, \theta_0, \theta_1, \theta_0\theta_1 \big \rangle_{\mathcal{O}_{\mathbb{P}^{1} }(U_z)},
\end{eqnarray}
where in the last equality we are conceiving it as a locally free $\mathcal{O}_{\mathbb{P}^{1} }$-module. Considering the transformation rules in the intersection, we find the following factorisation as $\mathcal{O}_{\mathbb{P}^{1} }$-modules:
\begin{eqnarray}
\mathcal{O}_{\mathbb{P}^{1|2} } = \mathcal{O}_{\mathbb{P}^{1} } \oplus \mathcal{O}_{\mathbb{P}^{1} } (-1)^{\oplus 2} \oplus \mathcal{O}_{\mathbb{P}^{1} } (-2).
\end{eqnarray}
Notice that the cohomology can be readily computed as $h^0 (\mathbb{P}^{1|2}, \mathcal{O}_{\mathbb{P}^{1|2} }) = 1$ and \\ $h^1 (\mathbb{P}^{1|2} , \mathcal{O}_{\mathbb{P}^{1|2} }) = 1$.\newline
Using a notation due to Witten \cite{WittenNotes}, the Berezinian sheaf over $\mathbb{P}^{1|2} $ is locally generated by an element of the form
\begin{eqnarray}
\mathcal{B}\mathpzc{er}_{\mathbb{P}^{1|2} } (U_z)= \big \langle \lbrack dz | d\theta_0 , d\theta_2] \big \rangle_{\mathcal{O}_{\mathbb{P}^{1|2} } (U_z)} .
\end{eqnarray}
Under a coordinate transformation, call it $\Phi$, taking local coordinates $w | \phi_0, \phi_1$ to $z | \theta_0, \theta_1$ as above, the Berezinian transforms as
\begin{eqnarray}
[dw | d\phi_0 , d \phi_1 ] \longmapsto [dz | d\theta_0 , d\theta_1] = \mbox{Ber}(\Phi)[dw | d\phi_0 , d \phi_1 ].
\end{eqnarray}
Therefore one gets:
\begin{eqnarray}
\mbox{Ber} \left [ \left (
\begin{array}{ccc}
- 1 / w^2 & 0 & 0 \\
- \theta_0 / w^2 & 1/ w & 0 \\
- \theta_1 / w^2 & 0 & 1/w
\end{array}
\right ) \right ] = \frac{-1/w^2}{1/w \cdot 1/w} = -1.
\end{eqnarray}
This trivial transformation implies the triviality. More precisely, viewing $\mathcal{B}\mathpzc{er}_{\mathbb{P}^{1|2} }$ as a $\mathcal{O}_{\mathbb{P}^{1} }$-module, one finds the following factorisation:
\begin{eqnarray}
\mathcal{B}\mathpzc{er}_{\mathbb{P}^{1|2} } \cong \mathcal{O}_{\mathbb{P}^{1} } \oplus \mathcal{O}_{\mathbb{P}^{1} } (-1)^{\oplus 2} \oplus \mathcal{O}_{\mathbb{P}^{1} } (-2) \cong \mathcal{O}_{\mathbb{P}^{1|2} },
\end{eqnarray}
and under the correspondence $1 \mapsto [dz | d\theta_0, d\theta_1]$, one has
\begin{eqnarray}
\mathcal{B}\mathpzc{er}_{\mathbb{P}^{1|2} } \cong \mathcal{O}_{\mathbb{P}^{1|2} },
\end{eqnarray}
as expected. The cohomology is obviously the same as the one of the structure sheaf.\newline
One can reach the same conclusions  for the structure sheaf and the Berezinian sheaf of $\mathbb{WP} _{(2)}^{1|1}$, remembering that in such a case one has transformations of the following form
\begin{eqnarray}
\left \{
\begin{array}{lcll}
w \longrightarrow z = \frac{1}{w} &  &  &  \\
\phi \longrightarrow \theta = \frac{\phi}{w^2}. &  &  &
\end{array}
\right.
\end{eqnarray}
Again, one finds that the Berezinian has a trivial transformation on the intersection and we have a correspondence $1 \mapsto [dz |d\theta] $ and an isomorphism
\begin{eqnarray}
\mathcal{B}\mathpzc{er}_{\mathbb{WP} ^{1|1}_{(2)}} \cong \mathcal{O}_{\mathbb{WP} ^{1|1}_{(2)}}.
\end{eqnarray}
This confirm that also the weighted projective space $\mathbb{WP}^{1|1}_{(2)}$ is a SCY variety, in the weak sense.


\subsection{The Sheaf Cohomology of Differential and Integral Forms}

To a large extent, generalisation of the ordinary commuting geometry to the richer context of supergeometry is pretty straightforward and it boils down to an application of the ``rule of sign'' \cite{Deligne} \cite{Manin}. One
issue stands out for its peculiarity: the theory of differential forms and integration. The issues related to this topic have been recently investigated by Catenacci et al. in a series of papers (here we will
particularly refer to \cite{Catenacci}) and reviewed by Witten in \cite{WittenNotes}. Below, we briefly sketch the main points, addressing the reader to the literature for details of the constructions. \newline
As one tries to generalise the complex of forms $(\Omega^{\bullet}, d^{\bullet} )$ to supergeometry using the $1$-superforms $\{d\theta^i\}_{i \in I}$ constructed out of the $\theta^i$, then it comes natural to define
wedge products such as $d\theta^1 \wedge \ldots \wedge d\theta^n$ to be \emph{commutative} in the $d\theta$'s, since the $\theta$'s are odd elements. This bears a very interesting consequence: the complex of
superforms $(\Omega_s^\bullet, d_s^\bullet)$ is bounded from below, but not from above! For example, superforms such as $(d\theta^i)^n \defeq d\theta^i \wedge \ldots \wedge d\theta^i$ do make sense and they are not zero, such as
their bosonic counterparts. The troublesome point is that there is no notion of a {top-form}, therefore a coherent notion of ``super integration'' is obtained only at the cost of enlarging the complex of
superforms and supplementing it with the so-called \emph{integral forms}. Using the notation of \cite{Catenacci}, the basic integral form are given by $\{\delta (d\theta^i)\}_{i \in I}$ and its higher derivatives $\{\delta^{(n)} (d\theta^i) \}_{i \in I}$, for $n>0$ (the \emph{degree of the integral form}). Here the use of the symbol $\delta$ should remind the Dirac delta distribution - and indeed an integral form satisfies similar properties \cite{Catenacci} -: it sets to zero terms in $d\theta^i$ and therefore, in some sense, it lowers the degree
of a superform. For this reason, an integral form is assigned a \emph{non-positive degree}: in fact, in the context of supergeometry one can also have forms with a negative degree. This fact can be better understood considering an example.
Let us take the super space $\mathbb{C} ^{2|2}$, and consider the following superform
\begin{eqnarray}
\omega_s = dz_1 dz_2 (d\theta^2)^4 \delta^{(2)} (d\theta^1),
\end{eqnarray}
where the wedge products are understood. Then $dz_1 dz_2 (d\theta^2)^4$ carries a degree of $6$, while the integral form $\delta^{(2)} (d\theta^1)$ lower the degree by $2$, so as a whole, we say that $\omega_s $ has degree $4 $
and we signal the presence of an integral form (of any degree) by saying that it has \emph{picture number} equal to $1$. Therefore, this enlarged complex of superforms is characterised by two numbers, the degree of the
form $n$ and their picture number $s$. The degree of the superform is the homogeneity degree in the differentials lowered by the total degree of the integral forms appearing in the monomial, whereas the picture number is the total number of the integral forms appearing in the monomial, regardless their degree. For example, we have that $\omega_s \in \Omega^{n=4;s=1}_{\mathbb{C} ^{2|2}}$. Notice, incidentally, that the picture number cannot exceed the odd dimension of the supermanifold and
operators linking complexes having different picture numbers - called \emph{picture changing operators} - can be defined. \vspace{4pt}\newline

In \cite{Catenacci}, the sheaf cohomology of superforms and integral forms of $\mathbb{P}^{1|1} $ has been studied, proving that just by adding an anti-commuting dimension, the cohomology becomes far richer. There is,
though, a substantial hole in the literature: no sheaf cohomology of superforms and integral forms has ever been computed for supermanifolds having extended supersymmetry, that is more than one odd dimensions. In
this scenario the computation of the cohomology for the case of $\mathbb{P}^{1|2} $ acquires value, besides being an example of cohomology of a SCY variety. \newline
We will see indeed that as soon as one has more than a single odd dimension, when the picture number is \emph{middle-dimensional} (that is, it is non-zero and not equal to the odd dimension of the manifold), then one finds
that the space of superforms is infinitely generated and its cohomology may be infinite-dimensional! \newline
This result calls, from a mathematical perspective, for a better understanding of the (algebraic) geometry of the complex of superforms and integral forms. Moreover, on the physical side, the possible usage and purposes of forms
having middle-dimensional picture number should be investigated and clarified. \vspace{4pt}

We now proceed to compute the sheaf cohomology of superforms of $\mathbb{P}^{1|2} $. In order to elucidate our method, we will carry out the computation in some details for the first case, namely for the space of superforms having null picture number, $\Omega_{\mathbb{P}^{1|2} }^{n;0}$; we then leave to the reader all the other cases, that follow the same pattern.\newline
As a $\mathcal{O}_{\mathbb{P}^{1|2} }$-module, $\Omega^{n;0}_{\mathbb{P}^{1|2} }$ is locally generated by:
\begin{eqnarray}
\Omega^{n;0}_{\mathbb{P}^{1|2} } (U_z) = \big \langle \{d\theta^i_0 d\theta^{n-i}_1 \}_{i = 0, \ldots, n}, \{dz d\theta_0^{j}
d\theta^{n-1-j}_{1} \}_{j=0, \ldots, n-1} \big \rangle_{\mathcal{O}_{\mathbb{P}^{1|2} } (U_z)}.
\end{eqnarray}
By looking at it as a (locally free) $\mathcal{O}_{\mathbb{P}^{1} }$-module, we can find the transformations of its generators. The first block of generators transform as (up to unimportant constants and signs)
\begin{align*}
& d\theta^i_0 d \theta_1^{n-i} = \frac{1}{w^n} d\phi_0^i d\phi_i^{n-i} +\frac{1}{w^{n+1}} \phi_0 dw d\phi_0^{i-1} d \phi_1^{n-i} + \frac{1}{w^{n+1}} \phi_1 dw d \phi_0^{n} d\phi_1^{n-1-i}, \\
& \theta_0 d\theta_0^i d\theta_1^{n-i} = \frac{1}{w^{n+1}} \phi_0 d\phi_0^i d \phi_1^{n-i} + \frac{1}{w^{n+2}} \phi_0 \phi_1 dw d\phi_0^i d\phi_1^{n-1-i}, \\
& \theta_1 d\theta_0^i d\theta_1^{n-i} = \frac{1}{w^{n+1}} \phi_1 d\phi_0^i d \phi_1^{n-i} + \frac{1}{w^{n+2}} \phi_0 \phi_1 dw d\phi_0^{i-1} d\phi_1^{n-i}, \\
& \theta_0 \theta_1 d\theta_0^i d\theta_1^{n-i} = \frac{1}{w^{n+2}} \phi_0 \phi_1 d\phi_0^i d \phi_1^{n-i}.
\end{align*}
The second block, instead, has only diagonal terms:
\begin{align*}
& dz d\theta_0^j d\theta^{n-1-j}_1 = \frac{1}{w^{n+1}}dw d\phi_0^i d \phi_1^{n-1-j}, \\
& \theta_0 dz d\theta_0^j d\theta^{n-1-j}_1 = \frac{1}{w^{n+2}}\phi_0 dw d\phi_0^i d \phi_1^{n-1-j}, \\
& \theta_1 dz d\theta_0^j d\theta^{n-1-j}_1 = \frac{1}{w^{n+2}}\phi_1 dw d\phi_0^i d \phi_1^{n-1-j}, \\
& \theta_0 \theta_1 dz d\theta_0^j d\theta^{n-1-j}_1 = \frac{1}{w^{n+2}} \phi_0 \phi_1 dw d\phi_0^i d \phi_1^{n-1-j},
\end{align*}
where we recall that $i = 0, \ldots, n$ and $j = 0, \ldots, n-1.$ Before proceeding further, we observe that, for $n$ fixed, looking at $\Omega^{n;0}_{\mathbb{P}^{1|2} }$ as a $\mathcal{O}_{\mathbb{P}^{1} }$-module, there are
\begin{eqnarray}
\dim_{\mathcal{O}_{\mathbb{P}^{1} }} {\Omega^{n;0}_{\mathbb{P}^{1|2} }} = 4 (n+1) + 4 n = 8 n + 4
\end{eqnarray}
terms in the factorisation; indeed, this is the dimension as a vector bundle/locally free sheaf of $\mathcal{O}_{\mathbb{P}^{1} }$-modules. \newline
We will pursue the strategy to group together pieces having
similar form, evaluating their transformations and afterwards factorising them into a direct sum of line bundles over $\mathbb{P}^{1} $ by means of Grothendieck's Theorem, by treating the off-diagonal terms in the
transition functions matrix: we recall that, in the notation above, we will be free to perform $\mathbb{C} [w]$-linear operations on the columns and $\mathbb{C} [1/w]$-linear operations on the rows. \newline
To this end, we now focus on the diagonal terms that do not need any further investigation: we will get $n+1$ terms and $n$ standing-alone terms out of $\theta_0 \theta_1 d\theta_0^i d\theta_1^{n-i}$
and $dz d\theta_0^j d\theta^{n-1-j}_1$, so these contribute to the factorisation with terms of the form
\begin{eqnarray}
\mathcal{O}_{\mathbb{P}^{1} } (-n-2)^{\oplus n+1} \oplus \mathcal{O}_{\mathbb{P}^{1} } (-n-1)^{\oplus n}.
\end{eqnarray}
So we are left with $8n+4 - (n +1) - n = 6n +3$ terms to give account to.
\newline
The other terms need some careful treatment. We start dealing with the terms coming from the transformation of $d\theta^i_0 d \theta_1^{n-i}$: these couples with the ones coming from $\theta_0 dz d\theta_0^j d\theta^{n-1-j}_1
$ and $\theta_1 dz d\theta_0^j d\theta^{n-1-j}_1$ whenever $i=j$ in the pairing with $\theta_0 dz d\theta_0^j d\theta^{n-1-j}_1$ and whenever $i = j+1$ in the pairing with $\theta_1 dz d\theta_0^j d\theta^{n-1-j}_1$. Since this holds true in the case $i=1, \ldots, n-1$, we need to consider $n-1$  identical $3 \times 3$ matrices of the following form:
\begin{eqnarray}
\left (
\begin{array}{ccc}
1/ w^{n} & 1/w^{n+1} & 1/w^{n+1} \\
0 & 1/w^{n+2} & 0 \\
0 & 0 & 1/w^{n+2}
\end{array}
\right ) \overset{C_1 - w C_2}{\longrightarrow} \left (
\begin{array}{ccc}
0 & 1/w^{n+1} & 1/w^{n+1} \\
-1/w^{n+1} & 1/w^{n+2} & 0 \\
0 & 0 & 1/w^{n+2}
\end{array}
\right ),  \notag
\end{eqnarray}
\begin{eqnarray}
\left (
\begin{array}{ccc}
0 & 1/w^{n+1} & 1/w^{n+1} \\
-1/w^{n+1} & 1/w^{n+2} & 0 \\
0 & 0 & 1/w^{n+2}
\end{array}
\right ) \overset{C_1 \leftrightarrow C_2}{\longrightarrow} \left (
\begin{array}{ccc}
1/w^{n+1} & 0 & 1/w^{n+1} \\
1/w^{n+2} & -1/w^{n+1} & 0 \\
0 & 0 & 1/w^{n+2}
\end{array}
\right ),  \notag
\end{eqnarray}
\begin{eqnarray}
\left (
\begin{array}{ccc}
1/w^{n+1} & 0 & 1/w^{n+1} \\
1/w^{n+2} & -1/w^{n+1} & 0 \\
0 & 0 & 1/w^{n+2}
\end{array}
\right ) \overset{R_2 - 1/w R_1}{\longrightarrow} \left (
\begin{array}{ccc}
1/w^{n+1} & 0 & 1/w^{n+1} \\
0 & -1/w^{n+1} & -1/w^{n+2} \\
0 & 0 & 1/w^{n+2}
\end{array}
\right ),  \notag
\end{eqnarray}
\begin{eqnarray}
\left (
\begin{array}{ccc}
1/w^{n+1} & 0 & 1/w^{n+1} \\
0 & -1/w^{n+1} & -1/w^{n+2} \\
0 & 0 & 1/w^{n+2}
\end{array}
\right ) \overset{C_3 - C_1}{\longrightarrow} \left (
\begin{array}{ccc}
1/w^{n+1} & 0 & 0 \\
0 & -1/w^{n+1} & -1/w^{n+2} \\
0 & 0 & 1/w^{n+2}
\end{array}
\right ),  \notag
\end{eqnarray}
\begin{eqnarray}
\left (
\begin{array}{ccc}
1/w^{n+1} & 0 & 0 \\
0 & -1/w^{n+1} & -1/w^{n+2} \\
0 & 0 & 1/w^{n+2}
\end{array}
\right ) \overset{R_2 + R_3}{\longrightarrow} \left (
\begin{array}{ccc}
1/w^{n+1} & 0 & 0 \\
0 & -1/w^{n+1} & 0 \\
0 & 0 & 1/w^{n+2}
\end{array}
\right ).  \notag
\end{eqnarray}
So this bit contributes with terms of the following form:
\begin{eqnarray}
\mathcal{O}_{\mathbb{P}^{1} } (-n-1)^{\oplus 2n-2} \oplus \mathcal{O}_{\mathbb{P}^{1} } (-n-2)^{\oplus n-1}
\end{eqnarray}
to be added to the previous ones. This boils the number of the remaining pieces down to $6n + 3 - (3n-3) = 3n + 6$. \newline
It is here worth pointing out that we have not given account for some terms in the counting above yet: we need indeed to consider separately $4$ terms that group into two identical $2\times 2$ matrices. Indeed the term $i = 0$
of $d\theta^i_0 d \theta_1^{n-i}$, that is $d\theta^n_1$, couples to the term $j=0$ (which was left out from the counting above) into the term $\theta_1 dz d\theta_0^j d\theta^{n-1-j}_1$, that is $\theta_1 dz d\theta_1^{n-1}$. This gives a
$2\times 2$ matrix of the form:
\begin{eqnarray}
\left (
\begin{array}{cc}
1/ w^{n} & 1/w^{n+1} \\
0 & 1/w^{n+2}
\end{array}
\right ) \overset{C_1 - w C_2 }{\longrightarrow} \left (
\begin{array}{ccc}
0 & 1/w^{n+1} &  \\
-1/w^{n+1} & 1/w^{n+2} &
\end{array}
\right ),  \notag
\end{eqnarray}
\begin{eqnarray}
\left (
\begin{array}{ccc}
0 & 1/w^{n+1} &  \\
-1/w^{n+1} & 1/w^{n+2} &
\end{array}
\right ) \overset{C_1 \leftrightarrow C_2 }{\longrightarrow} \left (
\begin{array}{ccc}
1/w^{n+1} & 0 &  \\
1/w^{n+2} & -1/w^{n+1} &
\end{array}
\right ),  \notag
\end{eqnarray}
\begin{eqnarray}
\left (
\begin{array}{ccc}
1/w^{n+1} & 0 &  \\
1/w^{n+2} & -1/w^{n+1} &
\end{array}
\right ) \overset{R_2 - 1/w R_1 }{\longrightarrow} \left (
\begin{array}{ccc}
1/w^{n+1} & 0 &  \\
0 & -1/w^{n+1} &
\end{array}
\right ).  \notag
\end{eqnarray}
The very same holds true in the case $i= n$ for $d\theta^i_0 d\theta_1^{n-i} $, that is $d\theta^n_0$, and in the case $j=n $ for $\theta_0 dz d\theta_0^j d\theta^{n-1-j}_1$, that is $\theta_0 dz \theta^{n-1}_0.$ So, we
have a pair of identical contributions that sums up to the ones already accounted:
\begin{eqnarray}
\mathcal{O}_{\mathbb{P}^{1} } (-n-1)^{\oplus 4}.
\end{eqnarray}
 All in all, this adds up 4 terms to the counting above, leaving us with $3n+2$ terms to be still accounted for. \newline
 The terms $\theta_0 d\theta_0^i d\theta_1^{n-i} $ and $\theta_1 d\theta_0^i d\theta_1^{n-i}$ couple with the last term, $\theta_0 \theta_1 dz
d\theta_0^j d\theta^{n-1-j}_1$, in the cases $i = 0, \ldots n-1$ for $\theta_0 d\theta_0^i d\theta_1^{n-i}$ and in the cases $i= 1, \ldots, n$ for $\theta_1 d\theta_0^i d\theta_1^{n-i}$, for all $j$. Therefore we have $3n$ identical $3\times 3 $ matrices of the form:
\begin{eqnarray}
\left (
\begin{array}{ccc}
1/ w^{n+1} & 0 & 1/w^{n+2} \\
0 & 1/w^{n+1} & 1/w^{n+2} \\
0 & 0 & 1/w^{n+3}
\end{array}
\right ) \overset{R_2 - R_1}{\longrightarrow} \left (
\begin{array}{ccc}
1/ w^{n+1} & 0 & 1/w^{n+2} \\
-1/w^{n+1} & 1/w^{n+1} & 0 \\
0 & 0 & 1/w^{n+3}
\end{array}
\right ),  \notag
\end{eqnarray}
\begin{eqnarray}
\left (
\begin{array}{ccc}
1/ w^{n+1} & 0 & 1/w^{n+2} \\
-1/w^{n+1} & 1/w^{n+1} & 0 \\
0 & 0 & 1/w^{n+3}
\end{array}
\right ) \overset{C_1 - C_2}{\longrightarrow} \left (
\begin{array}{ccc}
1/ w^{n+1} & 0 & 1/w^{n+2} \\
0 & 1/w^{n+1} & 0 \\
0 & 0 & 1/w^{n+3}
\end{array}
\right ),  \notag
\end{eqnarray}
\begin{eqnarray}
\left (
\begin{array}{ccc}
1/ w^{n+1} & 0 & 1/w^{n+2} \\
0 & 1/w^{n+1} & 0 \\
0 & 0 & 1/w^{n+3}
\end{array}
\right ) \overset{C_1 - wC_3}{\longrightarrow} \left (
\begin{array}{ccc}
0 & 0 & 1/w^{n+2} \\
0 & 1/w^{n+1} & 0 \\
1/w^{n+2} & 0 & 1/w^{n+3}
\end{array}
\right ),  \notag
\end{eqnarray}
\begin{eqnarray}
\left (
\begin{array}{ccc}
0 & 0 & 1/w^{n+2} \\
0 & 1/w^{n+1} & 0 \\
1/w^{n+2} & 0 & 1/w^{n+3}
\end{array}
\right ) \overset{C_1 \leftrightarrow C_3}{\longrightarrow} \left (
\begin{array}{ccc}
1/w^{n+2} & 0 & 0 \\
0 & 1/w^{n+1} & 0 \\
1/w^{n+3} & 0 & 1/w^{n+2}
\end{array}
\right ),  \notag
\end{eqnarray}
\begin{eqnarray}
\left (
\begin{array}{ccc}
1/w^{n+2} & 0 & 0 \\
0 & 1/w^{n+1} & 0 \\
1/w^{n+3} & 0 & 1/w^{n+2}
\end{array}
\right ) \overset{R_1 - 1/w R_3}{\longrightarrow} \left (
\begin{array}{ccc}
1/w^{n+2} & 0 & 0 \\
0 & 1/w^{n+1} & 0 \\
0 & 0 & 1/w^{n+2}
\end{array}
\right ).  \notag
\end{eqnarray}
Consequently, we have the following contribution to the factorisation:
\begin{eqnarray}
\mathcal{O}_{\mathbb{P}^{1} } (-n-1)^{\oplus n} \oplus \mathcal{O}_{\mathbb{P}^{1} } (-n-2)^{\oplus 2n}.
\end{eqnarray}
Notice that we have to take into account separately the terms corresponding to $i=n$ for $\theta_0 d\theta_0^i d\theta_1^{n-i}$ and to $i = 0$ for $\theta_1 d\theta_0^i d\theta_1^{n-i}$, yielding an identical (diagonal)
contribution of the form $\mathcal{O}_{\proj 1} (-n -1)^{\oplus 2}$.\newline
These last $2n +2$ terms complete the enumeration. Summing it all up, we are therefore ready to write down the whole factorisation for $n > 0$:
\begin{eqnarray}
\Omega^{n;0}_{\mathbb{P}^{1|2} } \cong \mathcal{O}_{\mathbb{P}^{1} } (-n-2)^{\oplus 4n} \oplus \mathcal{O}_{\mathbb{P}^{1} } (-n-1)^{\oplus 4n+4}.  \notag
\end{eqnarray}
Finally, we can count the dimensions of the cohomology groups:
\begin{eqnarray}
h^0 (\Omega^{n;0}_{\mathbb{P}^{1|2} }) = 0, \qquad h^1 (\Omega^{n;0}_{\mathbb{P}^{1|2} }) = 8n^2+8n.
\end{eqnarray}
This terminates the discussion of the differential forms with null picture number. All the other cases, having non-null picture number are treated in an analogous way, by remembering the transformation of the integral forms of type
$\delta^{(n)}(d\theta^i)$ \cite{Catenacci}. Below, we list their factorisation as $\mathcal{O}_{\mathbb{P}^{1} }$-modules. \newline
The space of superforms having maximal picture number and degree is locally generated by
\begin{eqnarray}
\Omega_{\mathbb{P}^{1|2} }^{1;2} (U_z) = \big \langle dz \delta (d\theta_0) \delta (d\theta_1) \big \rangle_{\mathcal{O}_{\mathbb{P}^{1|2} } (U_z)}.
\end{eqnarray}
Its transformation among the two charts yields the factorisation
\begin{eqnarray}
\Omega_{\mathbb{P}^{1|2} }^{1;2} \cong \mathcal{O}_{\mathbb{P}^{1} } \oplus \mathcal{O}_{\mathbb{P}^{1} } (-1)^{\oplus 2} \oplus \mathcal{O}_{\mathbb{P}^{1} } (-2) \cong \mathcal{O}_{\mathbb{P}^{1|2} }.
\end{eqnarray}
Thus, we can easily compute the dimensions of the cohomology groups, which are exactly the same as the ones of the structural sheaf:
\begin{eqnarray}
h^0 (\Omega_{\mathbb{P}^{1|2} }^{1;2}) = 1, \qquad h^1 (\Omega_{\mathbb{P}^{1|2} }^{1;2}) = 1.
\end{eqnarray}
It is not surprising that this is the same as the Berezinian line bundle over $\mathbb{P}^{1|2} $: indeed, elements of this sheaf are in some sense the supersymmetric analogues of the ordinary top-form for a manifold, and we
(Berezin-)integrate them, as one can integrate sections of the Berezinian sheaf. These two peculiar supersymmetric sheaves are fundamental in theory of integration on supermanifolds.\newline
We are left with the group $\Omega^{-n; 2}_{\mathbb{P}^{1|2} }$, for $n\geq 0 $ (which deserve some attention and bookkeeping such as $\Omega^{n; 0}_{\mathbb{P}^{1|2} }$). It is locally generated by
\begin{eqnarray}
\Omega^{-n;2}_{\mathbb{P}^{1|2} } (U_z) = \big \langle \{ \delta^{(i)} (d\theta_0) \delta^{(n-i)} (d\theta_1)\}_{i=0, \ldots, n}, d z\{
\delta^{(j)} (d\theta_0) \delta^{(n+1-j)} (d\theta_1)\}_{j=0, \ldots, n+1} \big \rangle_{\mathcal{O}_{\mathbb{P}^{1|2} } (U_z)},  \notag
\end{eqnarray}
which give the following factorisation
\begin{eqnarray}
\Omega^{-n;2}_{\mathbb{P}^{1|2} } \cong \mathcal{O}_{\mathbb{P}^{1} } (n+1)^{\oplus 4n+5} \oplus \mathcal{O}_{\mathbb{P}^{1} } (n)^{\oplus 4n+6} \oplus \mathcal{O}_{\mathbb{P}^{1} } (n-1).  \notag
\end{eqnarray}
The dimensions of the cohomology groups then read
\begin{eqnarray}
h^0 (\Omega^{-n;2}_{\mathbb{P}^{1|2} }) = 8 (n+2)(n+1), \qquad h^1 (\Omega^{-n;2}_{\mathbb{P}^{1|2} }) = 0.
\end{eqnarray}
Summarising, we have the following results for all cohomologies :
\begin{eqnarray}
h^0 (\Omega^{n;m}_{\mathbb{P}^{1|2} }) = \left \{
\begin{array}{lll}
1 &  & n=0,\; m=0 \\
0 &  & n>0, m=0 \\
8(n + 2)(n+1) &  & n\leq 0, \; m=2 \\
1 &  & n=1, \; m=2
\end{array}
\right. ,
\end{eqnarray}
\begin{eqnarray}
h^1 (\Omega^{n;m}_{\mathbb{P}^{1|2} }) = \left \{
\begin{array}{lll}
1 &  & n=0,\; m=0 \\
8 n (n + 1) &  & n>0, m=0 \\
0 &  & n \leq 0,m=2 \\
1 &  & n=1, \; m=2
\end{array}
\right. .
\end{eqnarray}
Notice that so far we have not carried out the computation of superforms having picture number equal to $1$: as anticipated, these are infinitely generated as a locally free sheaf, and they give infinite dimensional cohomology. \newline
The generators read
\begin{align}
\Omega_{\mathbb{P}^{1|2} }^{n \geq 0;1} (U_z) = & \big \langle \{ \delta^{(i)} (d \theta_0) d\theta_1^{n+i} \}_{i \in \mathbb{N} } , \; dz \{ \delta^{(i+1)}(d\theta_0) d\theta^{n+i}_1 \}_{i \in \mathbb{N} },  \notag \\
& \{ \delta^{(i)} (d \theta_1) d\theta_0^{n+i} \}_{i \in \mathbb{N} } , \; dz \{ \delta^{(i+1)}(d\theta_1) d\theta^{n+i}_0 \}_{i \in \mathbb{N} } \big \rangle_{\mathcal{O}_{\mathbb{P}^{1|2} (U_z)}}.
\end{align}
This is factorised as
\begin{eqnarray}
\bigoplus_{i \in \mathbb{N} } \left ( \mathcal{O}_{\mathbb{P}^{1} }^{\oplus 8} (-n-1) \oplus \mathcal{O}_{\mathbb{P}^{1} }^{\oplus 8}(-n) \right ),
\end{eqnarray}
and it yields $h^0 (\Omega_{\mathbb{P}^{1|2}}^{n > 0; 1})= 0$, while, remarkably, $h^0 (\Omega_{\mathbb{P}^{1|2}}^{0; 1})= h^1 (\Omega_{\mathbb{P}^{1|2} }^{n\geq 0; 1}) =\infty !$\\
Similarly, one finds
\begin{align}
\Omega_{\mathbb{P}^{1|2} }^{n < 0;1} (U_z) = & \big \langle \{ \delta^{(|n|+i)} (d \theta_0) d\theta_1^{i} \}_{i \in \mathbb{N} } , \; dz
\{ \delta^{(|n|+i+1)}(d\theta_0) d\theta^{i}_1 \}_{i \in \mathbb{N} }, \notag \\
& \{ \delta^{(|n|+i)} (d \theta_1) d\theta_0^{i} \}_{i \in \mathbb{N} } , \; dz \{ \delta^{(|n|+i+1)}(d\theta_1) d\theta^{i}_0 \}_{i \in \mathbb{N} } \big \rangle_{\mathcal{O}_{\mathbb{P}^{1|2} (U_z)}},
\end{align}
having factorisation
\begin{eqnarray}
\bigoplus_{i \in \mathbb{N} } \left ( \mathcal{O}_{\mathbb{P}^{1} }^{\oplus 8} (|n|-1) \oplus \mathcal{O}_{\mathbb{P}^{1} }^{\oplus 8}(|n|) \right )
\end{eqnarray}
which again gives infinite dimensional cohomology. \vspace{4pt}\newline

The computation of the cohomology of $\mathbb{WP} ^{1|1}_{(2)}$ is much easier, and it can be performed following the same lines as above. Also, having no middle picture number, there are no infinitely generated modules,
and no infinite cohomologies. \newline
By means of Grothendieck's Theorem, the complete sheaf cohomology can be computed to read
\begin{eqnarray}
h^0 (\Omega^{n;m}_{\mathbb{WP} ^{1|1}_{(2)}}) = \left \{
\begin{array}{lll}
1 &  & n=0,\; m=0 \\
0 &  & n>0, m=0 \\
4n+6 &  & n\leq 0, \; m=1 \\
1 &  & n=1, \; m=1
\end{array}
\right. ,
\end{eqnarray}
\begin{eqnarray}
h^1 (\Omega^{n;m}_{\mathbb{WP} ^{1|1}_{(2)}}) = \left \{
\begin{array}{lll}
1 &  & n=0,\; m=0 \\
8n &  & n>0, m=0 \\
0 &  & n\leq 0, \; m=1 \\
0 &  & n=1, \; m=1
\end{array}
\right. .
\end{eqnarray}
By the way, we signal a pathology, which looks like it may apply to any weighted projective space.
While the Berezinian sheaf is isomorphic to the structural sheaf of $\mathbb{WP} ^{1|1}_{(2)}$ - and indeed it has analogous factorisation and cohomology - one finds instead that the sheaf of the ``top-superform''
$\Omega_{\mathbb{WP} ^{1|1}_{(2)}}^{1;1} $ is not! To see this, it is enough to check the different factorisation and therefore the different cohomologies: one
finds indeed that $h^1 (\Omega^{0;0}_{\mathbb{WP} ^{1|1}_{(2)}}) = 1$ while $ h^1 (\Omega^{1;1}_{\mathbb{WP} ^{1|1}_{(2)}}) = 0.$


\subsection{de Rham Cohomology of $\pmb {\mathbb{WP}_{(2)}^{1|1}}$ and $ \pmb{\mathbb{P}^{1|2}}$}
 After calculating the sheaf cohomology of superforms on the super varieties $\mathbb{WP}_{(2)}^{1|1}$ and $\proj {1|2}$, we now aim at computing their \emph{holomorphic} de Rham cohomology. \\
Before we start, a remark on the adopted notation is due: given a supermanifold $\mathpzc{M}$, we will denote its de Rham cohomology groups as $H^{n;m}_{dR} (\mathpzc{M})$, where $n$ refers to the usual degree of the forms and $m$ refers to their picture number.  \\
We also stress that the boundary operator of the complex acts as $d : \mathcal{A}_\mathpzc{M}^{n; m} \rightarrow \mathcal{A}_\mathpzc{M}^{n+1; m},$ where $\mathcal{A}_{\mathpzc{M}}^{n;m}$ is the freely generated module of the $n$-forms having \emph{fixed} picture number $m$ that
are defined \emph{everywhere}, that is $\mathcal{A}_{\mathpzc{M}}^{n; m} \cong H^0 (\Omega_\mathpzc{M}^{n;m})$. In other words, the boundary operator $d$ does \emph{not} change the picture number of the form, and it just raises the degree of the form, so - as in
ordinary, purely bosonic geometry - we are just moving \emph{horizontally} on the complex, and we cannot jump from one complex to the other, by picture changing procedure.\vspace{4pt}\\

For clarity's sake, we start from the end of the treatment of the previous subsection, and we compute the de Rham cohomology of the weighted projective super space $\mathbb{WP}^{1|1}_{(2)}$, whose sheaf cohomology of superforms is always finite. We will
adopt a cumbersome but effective method, that has the advantages to display explicitly a basis of generators for the various de Rham groups. This is remarkable, for it possibly sets a more concrete ground for the observations in \cite{Bei-1} and especially
in \cite{Bei-2}, where it is observed that the \emph{BRST cohomology} of a (super) A-model is isomorphic to the cohomology of the superforms on the target space, that is on a supermanifold $\mathpzc{M}.$ \\
The starting point to compute the de Rham cohomology of the weighted projective super space $\mathbb{WP}^{1|1}_{(2)}$ is to look at its zeroth \v{C}ech cohomology, computed above. This actually yields two immediate results. The first one is that $H^{0;0}_{dR}
(\mathbb{WP}_{(2)}^{1|1}) = \mathbb{C}$, and it is generated by the constant function $1$. The second result that can be easily obtained is that $H^{n;0}_{dR} (\mathbb{WP}^{1|1}_{(2)}) = 0 $ for $n>0$; in fact, \v{C}ech cohomology guarantees that there
are no everywhere defined forms of degree $n>0$.\\
Let us now consider $H^{1;1}_{dR} (\mathbb{WP}_{(2)}^{1|1})$. \v{C}ech cohomology states that there is one everywhere defined form, locally given by $dz \delta^{(0)} (d\theta)$, that generates the group: this is also trivially \emph{closed} because in
particular $d (\delta^{(n)}(d\theta))  = 0$ (see \cite{Catenacci}). Thus, the natural question arises whether $dz \delta^{(0)} (d\theta) $ is exact or not. To answer this, one needs to look at $H^0 (\Omega^{0;1}_{\mathbb{WP}^{1|1}_{(2)}}) = \mathbb{C}^{6}$. In this case,
it is easy to see that
\bear
dz \delta^{(0)} (d\theta) = d (z \delta^{(0)} (d\theta)),
\eear
where $z \delta^{(0)} (d\theta) \in H^0 (\Omega^{0;1}_{\mathbb{WP}^{1|1}_{(2)}})$, so the form is exact and one obtains $H^{1;1}_{dR} (\mathbb{WP}^{1|1}_{(2)}) = 0$.\\
We now focus onto the de Rham cohomology connected to $H^{n;1} (\mathbb{WP}^{1|1}_{(2)}) = \mathbb{C}^{4n + 6}$, where $n \leq 0$.
Let us consider a generic form expanded as it belongs to a freely-generated module over $\mathcal{O}_{\proj 1}$. We have that:
\begin{align}
\omega = & \left ( F_0 (z) + \theta F_1 (z) \right ) \delta^{(n)} (d\theta) + \left ( G_0 (z) + \theta G_1 (z) \right ) dz \delta^{(n+1)} (d\theta) \nonumber \\
& = \left ( w^{n+2} F_0 (1/w)  + \phi w^n F_1 (1/z) \right ) \delta^{(n)} (d\phi) + \nonumber \\
& + \left ( - w^{n+1} G_0 (1/w) + \phi \left ( - w^{n+1} F_0 (1/ w ) - w^{n-1} G_1 (1/w) \right ) \right )dw \delta^{(n+1)} (d\phi)
\end{align}
where $F_0, F_1, G_0, G_1$ are polynomials. By changing the coordinates to $U_w$, we find that the form remains everywhere defined if and only if $\deg F_0 = n+ 2, \deg F_1 = n, \deg G_0 = n+1, \deg G_1 = n$, where $G_1$ has the constraint that the coefficient of its highest degree monomial is equal to the coefficient of the highest degree of $F_0$, which indeed yields a total of $4n + 6 $ free parameters, as already computed above. So far, this is nothing but another method to find the zeroth-dimensional \v{C}ech cohomology, without using Grothendieck's Theorem, as done in \cite{Catenacci} for the case of $\mathbb{P}^{1|1}$. On one hand, this method is certainly not efficient - especially as one needs to deal with more than one fermionic dimension -; on the other hand, in the context of the de Rham cohomology, it should be stressed that it has the advantage to make explicit the basis of the zeroth cohomology group of everywhere defined forms. \\
However, we should point out that a careful analysis of the various pieces involved in the computation carried out by exploiting Grothendieck's Theorem would have led to the same result in term of the basis of the space of everywhere defined forms. In the present treatment, we opted for this more cumbersome method, as long as the computations are easy-to-follow. \\
The most interesting group is the zeroth one : we find that $\deg F_0 = 2, \deg F_1 = 0, \deg G_0 = 1, \deg G_1 = 1$, and explicitly it holds that
\begin{align}
& F_0 (z) = a z^2 + b z + c, \\
& F_1 (z) = d, \\
& G_0 (z) = e z+ f,\\
& G_1 (z) = -a.
\end{align}
Grouping together the terms having the same coefficients, we obtain the following basis:
\begin{align}
&a \rightarrow \delta^{(0)} (d\theta) - \theta dz \delta^{(1)}(d\theta),\\
&b \rightarrow z \delta^{(0)} (d\theta),\\
&c \rightarrow z^2 \delta^{(0)} (d\theta),\\
&d \rightarrow \theta \delta^{(0)} (d\theta), \\
&e \rightarrow dz \delta^{(1)} (d\theta),\\
&f \rightarrow z dz \delta^{(1)} (d\theta).
\end{align}
Thus, one can verify that the module of the closed forms is generated by
\bear
Z_{dR}^{0;1} (\mathbb{WP}_{(2)}^{1|1}) = \big \langle \theta \delta^{(0)} (d\theta) , dz \delta^{(1)} (d\theta),z  dz \delta^{(1)} (d\theta)\big \rangle_{\mathcal{O}_{\proj 1}}.
\eear
Actually, the forms $dz \delta^{(1)} (d\theta),z  dz \delta^{(1)} (d\theta)$ are easily seen to be exact; indeed :
\begin{align}
& dz \delta^{(1)} (d\theta) = d \left (z \delta^{(1)} (d\theta) \right ), \\
& z dz \delta^{(1)} (d\theta) = d \left (\frac{1}{2} z^2 \delta^{(1)} (d\theta) \right),
\end{align}
and both the forms on the right-hand sides are everywhere defined, that is they belong to $H^0 (\Omega^{-1;1}_{\mathbb{WP}_{(2)}^{1|1}})$. One can then conclude that $H_{dR}^{0;1} (\mathbb{WP}^{1|1}_{(2)}) = \mathbb{C}$, and the group is generated by the closed
form $\theta \delta^{(0)} (\theta)$. \\
Writing explicitly the forms, we can see that \emph{all} the other groups $H^{n;1}_{dR} (\mathbb{WP}^{1|1}_{(2)}) $ for $n>0$ are trivial: one finds that $Z_{dR}^{n;1} (\mathbb{WP}^{1|1}_{(2)})$ is actually non-zero - there are closed forms -, but
$Z_{dR}^{n;1} (\mathbb{WP}^{1|1}_{(2)}) \cong B_{dR}^{n;1} (\mathbb{WP}^{1|1}_{(2)} )$ -, namely that all closed forms are exact and do not contribute to the de Rham cohomology.
Summing it all up, we have:
\bear
h_{dR}^{n;m} ({\set{WP}^{1|1}_{(2)}}) = \left \{ \begin{array}{lll} 1 & & n=0,\; m=0  \\
0 & & n>0, m=0 \\
1 & &  n= 0, \; m=1 \\
0 & & n\neq 0, \; m=1.
\end{array}
\right.
\eear
We now proceed to consider the holomorphic de Rham cohomology of $\mathbb{P}^{1|2}$: again, the starting point will be to look at the forms defined everywhere. By recalling the results on the sheaf cohomology of superforms obtained above, we see that
$H^{0;0}_{dR} (\proj{1|2}) = \mathbb{C}$ and it is generated by the constant function $1$, and $H^{n;0}_{dR} (\proj{1|2}) = 0$ (indeed, there are no globally defined forms).\\
Let us now consider the case $n=1, m=2$ - corresponding, as observed, to a kind of top-form -: the relative group is locally generated by the superform $dz \delta^{(0)}(d\theta_0) \delta^{(0)} (d\theta_1)$, which extends globally: this is certainly closed and
moreover, one can easily see, it is exact, for $d (z \delta^{(0)}(d\theta_0) \delta^{(0)} (d\theta_1)) = dz \delta^{(0)}(d\theta_0) \delta^{(0)} (d\theta_1)$ and $z \delta^{(0)}(d\theta_0) \delta^{(0)} (d\theta_1) \in H^0 (\Omega_{\proj{1|2}}^{0; 2})$. This implies that
$H_{dR}^{1;2} (\proj{1|2}) = 0$.\\
Next, we consider the groups $H_{dR}^{n; 2} (\proj{1|2})$ for $n\leq 0$. The most interesting case is given by $H_{dR}^{0;2} (\proj{1|2})$: the relative \v{C}ech cohomology group  has dimension 16 and we will study it carefully. We should be considering forms of the kind
\begin{align}
\omega = & (F_0 (z) + F_1 (z)\theta_0 + F_2 (z) \theta_1 + F_3 (z) \theta_0 \theta_1 ) \delta^{(0)} (d\theta_0) \delta^{(0)} (d\theta_1) + \nonumber \\
& (G_0 (z) + G_1 (z)\theta_0 + G_2 (z) \theta_1 + G_3 (z) \theta_0 \theta_1 ) dz \delta^{(0)} (d\theta_1) \delta^{(1)} (d\theta_1) +\nonumber \\
& (H_0 (z) + H_1 (z)\theta_0 + H_2 (z) \theta_1 + H_3 (z) \theta_0 \theta_1 ) dz \delta^{(1)} (d\theta_1) \delta^{(0)} (d\theta_1),  \nonumber
\end{align}
where the $F$'s, $G$'s and $H$'s are all polynomials, whose degree is identified as above, by studying whenever the form remains defined everywhere under a change of local chart, say from $U_z$ to $U_w$.\\
There are $10$ closed forms:
\begin{align}
Z^{0;2}_{dR} (\proj{1|2}) = & \big \langle \delta^{(0)} (d\theta_0) \delta^{(0)} (d\theta_1), \theta_0 dz \delta^{(0)} (d\theta_0) \delta^{(1)} (d\theta_1),  \theta_1 dz \delta^{(1)} (d\theta_0) \delta^{(0)} (d\theta_2) ,   \nonumber \\
& \theta_0 \delta^{(0)} (d\theta_0) \delta^{(0)}
(d\theta_1) , \theta_0 \delta^{(0)} (d\theta_0) \delta^{(0)} (d\theta_1) , z dz \delta^{(0)} (d\theta_0) \delta^{(1)} (d\theta_1), \nonumber \\
&dz \delta^{(0)} (d\theta_0) \delta^{(1)} (d\theta_1), z dz \delta^{(1)} (d\theta_0) \delta^{(0)} (d\theta_1), dz \delta^{(1)} (d\theta_0) \delta^{(0)} (d\theta_1), \cr
& \theta_1 \theta_2 \delta^{(0)} (d\theta_0) \delta^{(0)} (d\theta_1) \big \rangle_{\mathcal{O}_{\proj 1}}. \nonumber
\end{align}
The unique closed form that it is not exact is $\theta_0 \theta_1 \delta^{(0)} (d\theta_0)\delta^{(0)} (d\theta_1)$, which is therefore a generator for the group $H^{0:2}_{dR} (\proj{1|2}) = \mathbb{C}$. \\
Indeed, considering for example the closed form $dz \delta^{(1)} (d\theta_0) \delta^{(0)} (d\theta_1)$, one has:
\bear
dz \delta^{(1)} (d\theta_0) \delta^{(0)} (d\theta_1) = d (- \theta_1 dz \delta^{(1)} (d\theta_0) \delta^{(1)} (d\theta_1)),
\eear
where $d\theta_1 \delta^{(1)} (d\theta_1) = - \delta^{(0)} (d\theta_1)$ has been used. \\
As in the case of the weighted projective super space, proceeding in the negative degree sector, one finds that the closed forms are \emph{all} exact, and we have $H^{n;2}_{dR} (\proj{1|2}) = 0$ for $n \leq - 1$. This is ultimately to be connected to the
dimension of the space $H^0 (\Omega^{n;2}_{\proj{1|2}})$ for $n\leq -1$, and in turn to the transformation properties of the integral forms, giving rise to a huge space of globally defined forms. \vspace{5pt}

We now consider the space of everywhere defined forms having picture number equal to 1, which is somehow the most sensitive one, because, as we have seen above, it yields to an infinite dimensional sheaf cohomology. Before proceeding further, we recall that
$\Omega_{\proj {1|2}}^{n \geq 0;1}$ is infinitely generated as a locally free sheaf, and its generators read
\begin{align}
\Omega_{\proj {1|2}}^{n \geq 0;1} (U_z) = & \big \langle \{ \delta^{(i)} (d \theta_0) d\theta_1^{n+i} \}_{i \in \set{N}} , \; dz \{ \delta^{(i+1)}(d\theta_0) d\theta^{n+i}_1 \}_{i \in \set{N}}, \nonumber \\
& \{ \delta^{(i)} (d \theta_1) d\theta_0^{n+i} \}_{i \in \set{N}} , \; dz \{ \delta^{(i+1)}(d\theta_1) d\theta^{n+i}_0 \}_{i \in \set{N}}  \big \rangle_{\mathcal{O}_{\proj {1|2} (U_z)}}.
\end{align}
The factorisation is
\bear
\bigoplus_{i \in \set{N}} \left ( \mathcal{O}_{\proj 1}^{\oplus 8} (-n-1) \oplus \mathcal{O}_{\proj 1}^{\oplus 8}(-n) \right ).
\eear
Firstly, we observe that there are \emph{no} globally defined forms for $n>0$. Thus, it follows that the de Rham cohomology is $H^{n>0;1}_{dR} (\proj {1|2}) = 0$. \\
Secondly, all the other modules, for $n\leq 0$, gives an infinite dimensional zeroth (\v{C}ech) cohomology group. \\
As usual, we start analysing the $n= 0$ module. Since the generators read
\begin{align}
\Omega_{\proj {1|2}}^{0 ;1} (U_z) = & \big \langle \{ \delta^{(i)} (d \theta_0) d\theta_1^{i} \}_{i \in \set{N}} , \; dz \{ \delta^{(i+1)}(d\theta_0) d\theta^{i}_1 \}_{i \in \set{N}}, \nonumber \\
& \{ \delta^{(i)} (d \theta_1) d\theta_0^{i} \}_{i \in \set{N}} , \; dz \{ \delta^{(i+1)}(d\theta_1) d\theta^{i}_0 \}_{i \in \set{N}}  \big \rangle_{\mathcal{O}_{\proj {1|2} (U_z)}},
\end{align}
we can just deal with the first two blocks, and the other ones are symmetric up to the exchange $\theta_0 \leftrightarrow \theta_1.$\\
For the sake of convenience, let us consider separately the case $i = 0 $ and $i > 0 $, for some attention is requested as one deals with $i =0$ in the transformations.  \\
In this case, $i = 0$, one has:
\begin{align}
\omega = &(F_0 (z) + \theta_0 F_1 (z) + \theta_1 F_2 (z) + \theta_0 \theta_1 F_3 (z))\delta^{(0)} (d \theta_0) + \\
&(G_0 (z) + \theta_0 G_1 (z) + \theta_1 G_2 (z) + \theta_0 \theta_1 G_3 (z)) dz \delta^{(1)}(d\theta_0).
\end{align}
From \v{C}ech cohomology computations, we expect 4 free parameters that yield:
\begin{align}
H^0 (\Omega_{\proj {1|2}}^{0;1})\lfloor_{i = 0 } = \big \langle & z \delta^{(0)} (d\theta_0) - dz \delta^{(1)} (d\theta_0),  \delta^{(0)} (d\theta_0), \theta_0 \delta^{(0)} (d \theta_0), dz \delta^{(1)} (d\theta_0) \big \rangle.
\end{align}
The last three forms are closed, but only $\theta_0 \delta^{(0)} (d\theta_0)$ is \emph{not} exact, indeed
\begin{align}
\delta^{(0)} (d\theta_0) = d (- \theta_0 \delta^{(1)} (d\theta_0)) \qquad dz \delta^{(1)} (d\theta_0) = d (z \delta^{(1)} (d\theta_0))
\end{align}
and $- \theta_0 \delta^{(1)} (d\theta_0), z \delta^{(1)} (d\theta_0)$ are globally defined.
Analogously, we have that $\theta_1 \delta^{(0)} (d\theta_1)$ is closed and not exact, therefore it is non-zero in the quotient.\\
In the case $ i \neq 0 $, one is led to consider the transformation of
\begin{align}
\omega= & \left (F_0 (z) + F_1 (z)\theta_0 + F_2 (z) \theta_1 + F_3 (z) \theta_0 \theta_1 \right ) \delta^{(i)} (d\theta_0) d\theta^i_1 + \\
& \left ( G_0 (z) + G_1 (z)\theta_0 + G_2 (z) \theta_1 + G_3 (z) \theta_0 \theta_1 \right ) dz \delta^{(i+1)} (d\theta_0) d\theta_1^{i}.
\end{align}
One has:
\begin{align}
H^0 (\Omega_{\proj {1|2}}^{0;1})\lfloor_{i \neq 0 } = \big \langle  & z \delta^{(i)} (d\theta_0) d\theta_1^i + \theta_1 dz \delta^{(i+1)}(d\theta_0) d\theta_1^{i} + \theta_2 dz \delta^{(i)}(d \theta_0)d\theta_1^{i-1},  \delta^{(i)} (d\theta_0) d\theta_1^i, \nonumber \\
& \theta_0 \delta^{(0)} (d \theta_0) d\theta_1^i + \theta_1 \delta^{(i-1)}(d\theta_0) d\theta_1^{i-1}, dz \delta^{(i+1)} (d\theta_0) d\theta_1^i \big \rangle.
\end{align}
It can then be seen that $ \delta^{(i)} (d\theta_0) d\theta_1^i$ and $dz \delta^{(i+1)} (d\theta_0) d\theta_1^i$ are closed forms for every $i\geq 1$, but they are also exact, because
\bear
\delta^{(i)} (d\theta_0) d\theta_1^i = d (- \theta_0 \delta^{(i+1)} (d\theta_0)d\theta_1^i), \qquad dz \delta^{(i+1)} (d\theta_0) d\theta_1^i = d (z\delta^{(i+1)} (d\theta_0) d\theta_1^i), \nonumber
\eear
so there is no contribution to the cohomology. \\
This applies to each $n<0$, so there are no closed and not exact forms, and the complete holomorphic de Rham cohomology of $\proj{1|2}$ reads
\bear
h_{dR}^{n;m} (\proj{1|2}) = \left \{ \begin{array}{lll} 1 & & n=0,\; m=0,  \\
0 & & n>0, m=0, \\
2 & &  n= 0, \; m=1, \\
0 & & n\neq 0, \; m=1, \\
1 & & n = 0, \; m= 2, \\
0 & & n \neq 0, \; m=2.
\end{array}
\right.
\eear
The generators of the non-trivial groups are
\begin{align}
&H^{0;0}_{dR} (\proj {1|2})  = \big \langle 1 \big \rangle_{\mathcal{O}_{\proj 1}},\\
& H^{0;1}_{dR} (\proj {1|2}) = \big \langle \theta_1 \delta^{(0)} (d\theta_1), \;  \theta_2 \delta^{(0)} (d\theta_2) \big \rangle_{\mathcal{O}_{\proj 1}}, \\
& H^{0;2}_{dR} (\proj {1|2}) = \big \langle \theta_1 \theta_2 \delta^{(0)} (d\theta_1) \delta^{(0)} (d\theta_2) \big \rangle_{\mathcal{O}_{\proj 1}}.
\end{align}
As anticipated above, this is an interesting result, showing that the infinite dimensionality of \v{C}ech cohomology is cured at the level of the de Rham cohomology, which is the relevant one for physical applications, since it is
connected to the physical observables and it enters the evaluation of correlation functions \cite{Bei-2}. We would expect this kind of behaviour to be a feature of supermanifolds with more than one fermionic dimension.

\subsection{The complete de Rham cohomology of $\pmb {\mathbb{P}^{n|m}} $}
For completeness' sake as well as for future reference, we write down the whole holomorphic and real de Rham cohomology for general projective superspaces $\mathbb P^{n|m}$. This can be computed by using the same tedious direct method as above
(see also \cite{Catenacci}). \\
In the holomorphic case, one gets (notice that for $j=0$, $i$ cannot be negative)
\bear
H_{dR}^{i;j} (\proj{n|m}) = \left \{ \begin{array}{lll}
\mathbb C^{m\choose{j}} & &  i= 0, \; j=0,\ldots,m, \\
0 & & i\neq 0, \; j=0,\ldots,m.
\end{array}
\right.
\eear
In the real case, one obtains instead
\bear
H_{dR}^{i;j} (\proj{n|m}) = \left \{ \begin{array}{lll} \mathbb R^{m\choose{j}} & &  i= 2k,\ k=0,\ldots,n, \; j=0,\ldots,m, \\
0 & & i=2k+1,\ k=0,\ldots, n-1, \; j=0,\ldots,m.
\end{array}
\right.
\eear
The generators in the holomorphic case are given by a straightforward generalisation of the case $\mathbb P^{1|2}$ displayed above. In the real case, they are
\begin{eqnarray}
\omega_{k,I_j}\defeq \wedge^k \omega_{FS} \otimes \bigwedge_{\ell\in I_j}\theta_\ell \delta(d\theta_\ell),
\end{eqnarray}
where $I_j\subseteq \{0,1,\ldots,m\}$ has cardinality $j$, and $\omega_{FS}$ is the ordinary Fubini-Study form.


\subsection{Automorphisms and Deformations of $ \pmb{\mathbb{P}^{1|m}}$}

The method developed for the computation of the cohomology of projective super spaces over $\mathbb{P}^{1} $ easily allows us to evaluate the cohomology of the super tangent space, as well. \newline
Calculating the super Jacobian of the change of coordinates, we get
\begin{align}
& \partial_z = - w^2 \partial_w + w \sum_{i = 1}^n \phi_i \partial_{\phi_i} \\
& \partial_{\theta_i} = w \partial_{\phi_i}
\end{align}
with $i = 1, \ldots m$.
The super tangent sheaf is locally generated by the following elements:
\begin{eqnarray}
\qquad \mathcal{T}_{\mathbb{P}^{1|n} }U_z = \left \langle \partial_z ,
\{\theta_{J} \partial_z \}_{J= (j_1,\ldots, j_m )},
\{\partial_{\theta_i}\}^{i = 1, \ldots, m} , \{ \theta_{J}
\partial_{\theta_i}\}^{i = 1, \ldots, m}_{J= ( j_1, \ldots, j_m )} \right
\rangle_{\mathcal{O}_{\mathbb{P}^{1} }(U_z)},
\end{eqnarray}
where $J = (j_1, \ldots, j_m)$ is a multi-index such that $|J| = 1, \ldots, m $ and $j_i = \{0, 1\}$. For example, we can have elements like this: $\theta_1 \theta_3 \partial_z = \theta_{J = (1, 0, 1, 0 \ldots, 0) }
\partial_z. $ Thus, the total number of generators is $(m+1)\cdot2^m $. \newline
These have the following transformation rules:
\begin{align}
& \partial_z = -w^2 \partial_w + w \sum_{i = 1}^m \phi_i \partial_{\phi_i} \notag \\
& \theta_J \partial_z = \left (\frac{1}{w}\right )^{|J|-1} \phi_J \left (-w \partial_w +  \sum_{i = 1}^m \phi_i \partial_{\phi_i} \right )  \notag \\
& \partial_{\theta_i} = w \partial_{\phi_i}  \notag \\
& \theta_J \partial_{\theta_i} =\left ( \frac{1}{w} \right )^{|J|-1} \phi_J
\partial_{\phi_i},
\end{align}
where we stress that, depending on $J$, many terms might be zero in the transformation of $\theta_J \partial_z$ (namely, all the terms in the sum over $i$ such that $i \in J$). \\
Using Grothendieck's Theorem as above, one can compute the zeroth cohomology group of the tangent sheaf, whose dimension is:
\bear
h^0 (\mathcal{T}_{\proj {1|m}}) = (m + 2 )^2 -1 + \delta_{m, 2} 
\eear
Notice that $ (m + 2 )^2 -1 $ is just the number of generators of the Lie algebra associated to the super group $PGL (2|m)$, which is the supersymmetric generalisation of the ordinary M\"obius group $PGL (2, \mathbb{C})$, the automorphisms group of the projective line $\proj 1$.\\
It is worth noticing the presence of the \virgolette correction'' $\delta_{n,2}$, which, incidentally, makes its very appearance in the case of the super CY variety $\proj{1|2}$. This correspond to the presence of a further global vector field, (locally) given by $\theta_1 \theta_2 \partial_z \in H^0 (\mathcal{T}_{\proj{1|2}})$, which clearly does not belong to $\mathfrak{sl} (2|2)$, the Lie algebra of $PGL (2|2)$, as already noticed in \cite{ManinNC} and more recently in \cite{FioresiAut}. \\
Integrating this global vector field, we get the \virgolette finite'' version of the automorphism $\psi : \proj {1|2} \rightarrow \proj {1|2}$, called a \virgolette bosonisation'' in physics; locally, it is given by:
\begin{align}
&\psi \lfloor_{U_z}: (z,\theta_1,\theta_2)\longmapsto (z+\theta_1\theta_2,\theta_1,\theta_2), \\
& \psi \lfloor_{U_w}: (w,\phi_1,\phi_2)\longmapsto (w-\phi_1\phi_2,\phi_1,\phi_2).
\end{align}
Before we go on, it is important to stress that among \emph{all} the projective super spaces $\proj {n|m}$ - not only among $\proj {1|m}$! -, the case of $\proj {1|2}$ represents, remarkably, a \emph{unique exception}: indeed, it is the only case in which the automorphism group is larger than ${PGL}(n+1|m, \mathbb{C})$,\footnote{the bosonic reduction of ${PGL}(n+1|m)$} unlike to what stated in \cite{FioresiAut}. For reduced dimension 1 this exception has been first observed in \cite{ManinNC}, page 41. \\
This and other issues will be the subject of a forthcoming paper, where different methods to compute the cohomology of projective super spaces in a more general setting will be introduced and discussed.\vspace{4pt}

As for the deformations, given by $h^1 (\mathcal{T}_{\proj{1|m}})$, one finds
\begin{align}
h^1 (\mathcal{T}_{\proj {1|m}}) = (m+2) \left [(m+2) + (m-4) 2^{m-1} \right ] - (m-2) 2^{m-1} - 1.
\end{align}
We can see therefore that $\proj {1|1}$, together with $\proj {1|3}$ and the super CY variety $\proj {1|2}$ are \emph{rigid} as they have no deformations, while in the case $m \geq 4$, we start finding a non-zero $h^1 (\mathcal{T}_{\proj {1|m}})$. For instance, for $m = 4$ we find $h^1 (\mathcal{T}_{\proj {1|4}}) = 19$. We leave to future works a careful investigation of the structure of these deformations.

\section{A Super Mirror Map for SCY in Reduced Dimension $\pmb 1$}

\label{sec:mirror map}

In \cite{Sethi} the conjecture has been put forward that the puzzle of mirror of rigid (ordinary) CY manifolds could be solved by enlarging the relevant category for mirror symmetry, including also super manifolds, in
particular SCY manifolds. Later on, triggered by previous studies in \cite{NeiVafa} and \cite{WittenTwistor}, Aganagic and Vafa proposed a path integral argument to obtain the mirror of Calabi-Yau supermanifolds as super
Landau-Ginzburg (LG) theories \cite{AgaVafa}: the construction is exploited to compute the mirror of SCY manifolds in toric varieties and in particular to compute the mirror of the ``twistorial'' (actually super) Calabi-Yau $\mathbb{P}^{3|4} $
\cite{WittenTwistor}. Remarkably, after a suitable limit of the K\"ahler parameter $t$, the mirror has a geometric interpretation: indeed, it is a quadric in the product space $\mathbb{P}^{3|3} \times \mathbb{P}^{3|3} $, and it is again a SCY manifold.

Since we are interested into enlarging the mirror symmetry map for elliptic curves to a supersymmetric context, here we will apply the construction of \cite{AgaVafa} to the case of bosonic dimension equal to $1$ and reduced manifold given by $\proj {1}$, i.e. to the
two SCY's $\mathbb{P}^{1|2} $ and $\mathbb{WP} _{(2)}^{1|1}$. In doing that, in contrast with \cite{AgaVafa}, we will not need to take any limit of the K\"ahler parameter: in fact, a further geometric investigation, carried out by some
suitable change of coordinates, shows that $\mathbb{P}^{1|2} $ is actually self-mirror and it is mapped to itself. The mirror of the weighted projective super space $\mathbb{WP} _{(2)}^{1|1}$ instead is not a geometry. \newline
Before proceeding to the actual computation, it should be here remarked that a further, mathematically oriented, analysis needs to be carried out. Despite the effort in \cite{AgaVafa}, many issues are still unsettled, as
for example the role of the K\"ahler parameter $t$. It is indeed a matter of question how to define, mathematically and in full generality, a super analogue of the ordinary K\"ahler condition, and
therefore how to identify a super K\"ahler variety.

\subsection{Mirror Construction for $\pmb {{\mathbb{P}^{1|2}}}$}

Following \cite{AgaVafa}, we construct the dual of the \emph{LG model} associated to $\mathbb{P}^{1|2}$: it turns out this is given by a \emph{$\sigma$-model} on a super Calabi-Yau variety in $\mathbb{P}^{1|1} \times
\mathbb{P}^{1|1} $, which is again a SCY variety given by $\mathbb{P}^{1|2} $. In other words, $\mathbb{P}^{1|2} $ gets mapped to itself! \newline
We will focus on the holomorphic part of the (super)potential, where $X_I, Y_I$ for $I=0,1$ are \emph{bosonic/even} super fields and $\eta_I, \chi_I$ for $I=0,1$ are \emph{fermionic/odd} super fields (i.e., the lowest component of
their expansion is a bosonic field and a fermionic field, respectively), while $t$ is the K\"ahler parameter, mentioned above. This is given by
\begin{align}
\mathcal{W}_{\mathbb{P}^{1|2} } (X, Y, \eta, \xi) = \int & \prod_{I=0}^1 \mathcal{D}Y_I\mathcal{D}X_I \mathcal{D}\eta_I \mathcal{D}\chi_I \delta
\left ( \sum_{I=0}^1 (Y_I - X_I) - t \right ) \cr
& \cdot \exp \left \{ \sum_{I=0}^1 e^{-Y_I} + e^{-X_I} + e^{-X_I}\eta_I\chi_I \right \}.  \notag
\end{align}
By a field redefinition,
\begin{eqnarray}
X_1 = \hat X_1 + Y_0, \qquad \qquad Y_1 = \hat Y_1 + Y_0,
\end{eqnarray}
the path-integral above can be recast as follows :
\begin{align}
\int & \mathcal{D}Y_0 \mathcal{D}X_0 \mathcal{D}\hat Y_1 \mathcal{D}\hat X_1
\prod_{I=0}^1 \mathcal{D}\eta_I \mathcal{D}\chi_I \delta \left ( Y_0 - X_0 +
Y_1 -X_1 -t \right )  \notag \\
& \cdot \exp \left \{ e^{-Y_0} + e^{-X_0} + e^{-\hat Y_1 - Y_0} + e^{- \hat
X_1 - Y_0} + e^{-X_0}\eta_0\chi_0 + \eta_1 \chi_1 e^{- \hat X_1 - Y_0 }
\right \}.  \notag
\end{align}
Integrating in $X_0$, the delta imposes the following constraint on the bosonic fields:
\begin{eqnarray}
X_0 = Y_0 + (Y_1 - X_1) -t.
\end{eqnarray}
Plugging this inside the previous path integral one gets
\begin{align}
\int \mathcal{D}Y_0 \mathcal{D}\hat Y_1 \mathcal{D}\hat X_1 \prod_{I=0}^1
\mathcal{D}\eta_I \mathcal{D}\chi_I & \exp \left \{ e^{-Y_0} + e^{-Y_0 - (Y_1 -X_1) + t} + e^{-\hat Y_1 - Y_0} +e^{- \hat X_1 - Y_0} \right\} \cr
&\cdot \exp \left\{ e^{-Y_0 - (Y_1 -X_1) + t }\eta_0\chi_0 + \eta_1 \chi_1 e^{- \hat X_1 - Y_0 } \right \}.  \notag
\end{align}
The fermionic $\mathcal{D}\eta_0 \mathcal{D}\chi_0$ integration reads
\begin{align}
&\int \mathcal{D}\eta_0 \mathcal{D}\chi_0 \exp \left \{ e^{-Y_0 - (Y_1 -X_1)+ t } \eta_0 \chi_0 \right \} =  \notag \\
& = \int \mathcal{D}\eta_0 \mathcal{D}\chi_0 e^{-Y_0 - (Y_1 -X_1) + t } \left ( 1 + \eta_0 \chi_0 \right ) = - e^{-Y_0 - (Y_1 -X_1) + t },
\end{align}
and therefore one obtains that
\begin{align}
- \int \mathcal{D}Y_0 &\mathcal{D}\hat Y_1 \mathcal{D}\hat X_1 \mathcal{D} \eta_1 \mathcal{D}\chi_1 e^{-Y_0 - (Y_1 -X_1) + t } \notag \\
&\cdot  \exp \left \{ e^{-Y_0} \left ( 1+ e^{- (Y_1 -X_1) + t} + e^{-\hat Y_1} + e^{- \hat X_1 } + \eta_1 \chi_1 e^{- \hat X_1 } \right ) \right \}.  \notag
\end{align}
$e^{-Y_0}$ might be interpreted as a multiplier, and we perform the coordinate charge
\begin{eqnarray}
e^{-Y_0} = \Lambda, \qquad \qquad \mathcal{D}Y_0 = - \Lambda^{-1} \mathcal{D} \Lambda,
\end{eqnarray}
such that the integral reads
\begin{align}
\int \Lambda^{-1 } \mathcal{D}\Lambda &\mathcal{D}\hat Y_1 \mathcal{D}\hat X_1 \mathcal{D}\eta_1 \mathcal{D}\chi_1 \Lambda e^{ - (Y_1 -X_1) + t }  \notag \\
& \cdot \exp \left \{ \Lambda \left ( 1+ e^{- (Y_1-X_1) + t} + e^{-\hat Y_1} + e^{- \hat X_1 } + \eta_1 \chi_1 e^{- \hat X_1 } \right ) \right \}.  \notag
\end{align}
Finally, by performing another field redefinition, namely
\begin{eqnarray}
e^{- \hat X_1} = x_1, \qquad \qquad \mathcal{D} \hat X_1 = - \frac{\mathcal{D}x_1}{x_1}, \\
e^{- \hat Y_1} = x_1 y_1, \qquad \qquad \mathcal{D} \hat Y_1 = - \frac{\mathcal{D}y_1}{y_1}, \\
\eta_1 = \frac{\tilde \eta_1}{x_1}, \qquad \qquad \mathcal{D}\eta = x_1
\mathcal{D}\tilde \eta,
\end{eqnarray}
we notice that the Berezinian enters the transformation of the measure! In fact, the path-integral acquires the following form :
\begin{align}
\mathcal{W}_{\mathbb{P}^{1|2} } & = \int \mathcal{D}\Lambda \frac{\mathcal{D} y_1}{y_1} \frac{\mathcal{D} x_1}{x_1} (x_1 \mathcal{D}\tilde \eta_1)
\mathcal{D}\chi_1 \left (y_1 e^{ t } \right ) \exp \left \{ \Lambda \left (1+ e^{t} y_1 + x_1+ x_1 y_1 + \tilde \eta_1 \chi_1 \right ) \right \}
\notag \\
& = \int \mathcal{D}\Lambda {\mathcal{D} y_1} {\mathcal{D} x_1} \mathcal{D}\tilde \eta_1 \mathcal{D}\chi_1 e^{t } \exp \left \{ \Lambda
\left ( 1+ e^{t} y_1 + x_1+ x_1 y_1 + \tilde \eta_1 \chi_1 \right ) \right\}.
\end{align}
By noticing that the factor $e^t$ is not integrated over, and performing the integration over the Lagrange multiplier $\Lambda$, one obtains that the theory is constrained on the hypersurface
\begin{eqnarray}
1 + x_1 + x_1 y_1 + \tilde \eta \chi + e^t y_1 = 0.
\end{eqnarray}
By redefining the field $\tilde y_1 = 1+ y_1$, a more symmetric form can be achieved :
\begin{eqnarray}
1 + x_1 \tilde y_1 + \tilde \eta \chi + e^t ( \tilde y_1 - 1) = 0.
\end{eqnarray}
Casting the equation in homogeneous form, we have
\begin{eqnarray}
\mathbb{P}^{1|1} \times \mathbb{P}^{1|1} \supset X_0 \tilde Y_0 + X_1 \tilde Y_1 + \tilde \eta \chi + e^t (X_0 \tilde Y_1 - X_0 \tilde Y_0) =0.
\end{eqnarray}
This is a quadric, call it $\mathcal{Q}$, in $\mathbb{P}^{1|1} \times \mathbb{P}^{1|1} $, with homogeneous coordinates $[X_0 : X_1 : \tilde \eta ]$ and $[\tilde Y_0 : \tilde Y_1 : \chi]$ respectively, and it is a super
Calabi-Yau manifold. In the following treatment, we will drop the tildes and we will just call the homogenous coordinates of the super projective spaces $[X_0 : X_1 :\eta] \equiv [X_0 : X_1 : \tilde \eta ]$ and $[Y_0 : Y_1 : \eta] \equiv [\tilde Y_0 : \tilde Y_1 : \chi].$
We now re-write the equation for $\mathcal{Q}$ in the following form:
\begin{eqnarray}
X_0 ((1-e^t)Y_0 + e^t Y_1) + X_1 Y_1 + \eta \chi = 0.
\end{eqnarray}
Setting
\begin{eqnarray}
\ell (Y_0 , Y_1) \defeq (1-e^t) Y_0 + e^tY_1,
\end{eqnarray}
it is not hard to see that the reduced part $\mathcal{Q}_{red}$ in $\mathbb{P}^{1 } \times \mathbb{P}^{1} $ is obtained just by setting the odd coordinates to zero, as
\begin{eqnarray}
\mathbb{P}^{1 } \times \mathbb{P}^{1 } \supset X_0 \, \ell (Y_0 , Y_1 ) + X_1 Y_1 = 0,
\end{eqnarray}
and one can realize that $\mathcal{Q}_{red} \cong \mathbb{P}^{1} $. \newline
We are interested into fully identifying $\mathcal{Q}$ as a known variety; to this end, we observe that, as embedded into $\mathbb{P}^{1|1} \times \mathbb{P}^{1|1} $, it is covered by the Cartesian product of the usual four open
sets:
\begin{align}
&U_0 \times V_0 = \{ [X_0 : X_1: \eta] : X_0 \neq 0\} \times \{ [Y_0 : Y_1 :
\chi] : Y_0 \neq 0\}, \\
&U_0 \times V_1 = \{ [X_0 : X_1: \eta] : X_0 \neq 0\} \times \{ [Y_0 : Y_1 :
\chi] : Y_1 \neq 0\}, \\
&U_1 \times V_0 = \{ [X_0 : X_1 : \eta ] : X_1 \neq 0\} \times \{ [Y_0 : Y_1
: \chi] : Y_0 \neq 0\}, \\
&U_1 \times V_1 = \{ [X_0 : X_1: \eta ] : X_1 \neq 0\} \times \{ [Y_0 : Y_1
: \chi] : Y_1 \neq 0\}.
\end{align}
Moreover, one needs all the above four open sets to cover $\mathcal{Q}$, because
\begin{align}
& \mathcal{Q}_{red} \cap \{ X_0 = 0 \} = [0:1] \times [1:0] \in U_1 \times
V_0, \\
& \mathcal{Q}_{red} \cap \{ X_1 = 0 \} = [1:0] \times [1:1-e^{-t}] \in U_0
\times V_0, \\
& \mathcal{Q}_{red} \cap \{ Y_0 = 0 \} = [1: -e^t] \times [0:1] \in U_0
\times V_1, \\
& \mathcal{Q}_{red} \cap \{ X_0=X_1 = 1 \} = [1: 1] \times [e^t+1:e^t-1] \in U_1 \times
V_1.
\end{align}
Therefore, we would like to find a suitable change of coordinates allowing us to use fewer open sets. It turns out that one can reduce to use only two open sets. Indeed, by switching coordinates to
\begin{align}
&Y^\prime_0 \defeq \ell (Y_0, Y_1) , \quad & Y^\prime_1 \defeq Y_1, \\
&X^\prime_0 \defeq X_0, \quad & X^\prime_1 \defeq X_1 , \\
&\eta^\prime \defeq \eta, \quad & \chi^\prime \defeq \chi,
\end{align}
the equation for $\mathcal{Q}$ becomes
\begin{eqnarray}
X^\prime_0 Y^\prime_0 + X^\prime_1 Y^\prime_1 + \eta^\prime \chi^\prime = 0.
\end{eqnarray}
Then, by exchanging $Y^\prime_0 $ with $Y^\prime_1$ and dropping the primes for convenience, one obtains the following equation for $\mathcal{Q}$ :
\begin{eqnarray}
X_0 Y_1 + X_1 Y_0 + \eta \chi = 0.
\end{eqnarray}
Since
\begin{align}
& \mathcal{Q}_{red} \cap \{X_0 = 0\} = \mathcal{Q}_{red} \cap \{Y_0 = 0\} =
[0:1] \times [0:1] \in U_1 \times V_1, \\
& \mathcal{Q}_{red} \cap \{X_1 = 0\} = \mathcal{Q}_{red} \cap \{Y_1 = 0\} =
[1:0] \times [1:0] \in U_0 \times V_0,
\end{align}
this change of coordinates allows us to cover $\mathcal{Q}$ by just two open sets, namely by :
\begin{align}
& U_\mathcal{Q} \defeq \mathcal{Q} \cap (U_0 \times V_0), \\
& V_{\mathcal{Q}} \defeq \mathcal{Q} \cap (U_1 \times V_1).
\end{align}
Therefore, by choosing the following (affine) coordinates:
\begin{align}
& U_\mathcal{Q} : \; z \defeq \frac{X_1}{X_0}, \quad u \defeq \frac{Y_1}{Y_0}, \quad \theta_{0} \defeq \frac{\eta}{X_0}, \quad \theta_1 \defeq \frac{\chi}{Y_0}, \\
& V_\mathcal{Q} : \; w \defeq \frac{X_0}{X_1}, \quad v \defeq \frac{Y_0}{Y_1}, \quad \phi_{0} \defeq -\frac{\eta}{X_1}, \quad \phi_1 \defeq \frac{\chi}{Y_1},
\end{align}
the following two affine equations for $\mathcal{Q}$ of $U_{\mathcal{Q}}$ and $V_{\mathcal{Q}}$ are respectively obtained :
\begin{align}
& U_\mathcal{Q} : \; z + u + \theta_0 \theta_1 = 0,\label{eq:UQ} \\
& V_\mathcal{Q} : \; w + v - \phi_0 \phi_1 = 0, \label{eq:VQ}
\end{align}
describing lines in $\mathbb{C}^{2|2}.$ We notice that these two equations are glued together using the relations
\begin{align}
& w = \frac{1}{z}, & v = \frac{1}{u}, \\
& \phi_0 = - w \theta_0, & \phi_1 = v \theta_1.
\end{align}
Finally, we would like to characterise the variety $\mathcal{Q}$ by its transition functions, in order to identify it with a known one. By the previous equation, we may take as \emph{proper} bosonic coordinates $u$ and $v$, as
\begin{align}
& z = - u - \theta_0 \theta_1, \\
& w = - v + \phi_0 \phi_1.
\end{align}
We already know that $v = \frac{1}{u}$ and $\phi_1 = \frac{\theta_1}{u} $, so we still have to deal with $\phi_0:$
\begin{align}
\phi_0 & = - \frac{\theta_0}{z} = \frac{\theta_0}{u + \theta_0 \theta_1} =
\frac{\theta_0 (u - \theta_0 \theta_1)}{(u + \theta_0 \theta_1 )(u - \theta_0
\theta_1 )} = \frac{\theta_0 u}{u^2} = \frac{\theta_0}{u},
\end{align}
implying that the variety $\mathcal{Q} \subset \mathbb{P}^{1} \times \mathbb{P}^{1} $ is actually nothing but $\mathbb{P}^{1|2} $.\newline
This shows that the super mirror map proposed by Vafa and Aganagic makes the supermanifold $\mathbb{P}^{1|2} $ self-mirror, actually it is mapped to itself. This goes along well with what holds for elliptic
curves: indeed, an elliptic curve is the mirror of another elliptic curve.


\subsection{$ \pmb{\mathbb{P}^{1|2}}$ as a $\pmb {N=2}$ Super Riemann Surface}

We recall that a $N=2$ super Riemann surface is, by definition, a $1|2$ complex supermanifold $M$ such that the super tangent sheaf $\mathcal{T}_M$ has two $0|1$ subbundles $\mathcal{D}_1$ and $\mathcal{D}_2$, locally generated by vector fields
$D_1,D_2$ that are integrable, i.e. $D_i^2=fD_i$ for some odd function, and $\mathcal{D}_1\otimes\mathcal{D}_2, \mathcal{D}_1,\mathcal{D}_2$ generate $\mathcal{T}_M$ at any point. We address the interested reader to \cite{FaRe-1990} and \cite{Witten-1} for details, as well as to the more recent articles \cite{Bei-1} and \cite{Bei-2} for further developments and some physical interpretations.

Below, we will show that $\mathbb{P}^{1|2}$ is indeed a $N=2$ super Riemann surface. In order to find the needed $0|1$ line bundles $\mathcal{D}_1$ and $\mathcal{D}_2$, we adopt the method proposed in \cite{Witten-1} at page 107, namely we will find two maps
$p_1:\mathbb{P}^{1|2}\to X_1$ and $p_2:\mathbb{P}^{1|2}\to X_2$, with $X_1,X_2$ two suitable $1|1$ supermanifolds, and we will define $\mathcal{D}_i$ as the sheaf kernel of the differential $dp_i:\mathcal{T}_{\mathbb{P}^{1|2}}\to p_i^\ast\mathcal{T}_{X_i}$.
These two maps can immediately be determined from the model of $\mathbb{P}^{1|2}$ contained in $\mathbb{P}^{1|1}\times \mathbb{P}^{1|1}$ found in the previous section, in which we computed the mirror of $\mathbb{P}^{1|2}$. Indeed, we can set
$X_1=X_2=\mathbb{P}^{1|1}$ and the map $p_i$ equal to the restriction of the $i$-th projection $\pi_i:\mathbb{P}^{1|1}\times \mathbb{P}^{1|1}\to \mathbb{P}^{1|1}$ to $\mathbb{P}^{1|2}$. In order to give explicit local calculations of the vector fields $D_1,D_2$
that generate the line bundles $\mathcal{D}_1,\mathcal{D}_2$ and to show that they have all the required properties, we can exploit Eqs. (\ref{eq:UQ}) and (\ref{eq:VQ}) of the open sets $U_\mathcal{Q}$ and $V_\mathcal{Q}$ as sub-supermanifolds of
the open affine $\mathbb{A}^{2|2}\subset \mathbb{P}^{1|1}\times \mathbb{P}^{1|1}$ with coordinates $z,u,\theta_0,\theta_1$. For example, from the equation $$z+u+\theta_0\theta_1=0$$
in $\mathbb{A}^{2|2}$, we see that
\begin{align}
& p_1(z,u,\theta_0,\theta_1)=(z,\theta_0)\\
& p_2(z,u,\theta_0,\theta_1)=(u,\theta_1).
\end{align}
Then, $\mathcal{D}_1$ has sections given by those vector fields $\alpha\partial_z+\beta\partial_u+\gamma\partial_{\theta_0}+\delta\partial_{\theta_1}$ that vanish on the elements
$ z,\ \theta_0,\ z+u+\theta_0\theta_1$. This implies $\alpha=\gamma=0$ and $\beta=\delta\theta_0$, and therefore they are multiples of $$D_1=\partial_{\theta_1}+\theta_0\partial_u.$$ Similarly, one finds that the vector field  $$D_2=\partial_{\theta_0}-\theta_1\partial_z$$
generates all the vector fields on $U_\mathcal{Q}$ that vanish on $u,\ \theta_1,\ z+u+\theta_0\theta_1$. Since $D_1$ and $D_2$ vanish on $z+u+\theta_0\theta_1$, they are tangent vector fields on $U_\mathcal{Q}$ that, by construction, generate the kernels
$\mathcal{D}_1$ and $\mathcal{D}_2$ of the differentials $dp_1$ and $dp_2$.
The reader can easily check that $D_1^2=D_2^2=0$, and that $$\{D_1,D_2\}=D_1D_2+D_2D_1=\partial_u-\partial_z;$$ moreover, this latter is equal to $\partial_u$ when evaluated on an element of $\mathcal{O}_{U_{\mathcal{Q}}}$.  Since $u$ is  a
bosonic coordinate for $U_\mathcal{Q}$, one ralizes that $\{D_1,D_2\}, D_1, D_2$ generate $\mathcal{T}_{\mathbb{P}^{1|2}}$ at any point of $U_\mathcal{Q}$. Similar formulas can be obtained for the open $V_\mathcal{Q}$.


\subsection{Mirror Construction for $ \pmb{\mathbb{WP} _{(2)}^{1|1}}$}

In the case of weighted projective super space, we need to evaluate the following (super)potential in order to find the dual theory:
\begin{align}
\mathcal{W}_{\mathbb{WCP} ^{1|1}_{(1,1 | 2)}} = \int (\mathcal{D} Y_1
\mathcal{D} Y_2 )\mathcal{D}X \mathcal{D}\eta \mathcal{D} \chi & \delta \left
( Y_1 + Y_2 - 2X - t\right ) \cr
&\cdot\exp \left \{ e^{-Y_1} + e^{-Y_2} + e^{-X} (1 + \eta \chi)\right \}.
\end{align}
Performing the integration in the fermionic variables, one obtains
\begin{align}
\mathcal{W}_{\mathbb{WCP} ^{1|1}_{(1,1 | 2)}} = \int (\mathcal{D} Y_1 \mathcal{D}
Y_2 )\mathcal{D}X e^{-X} & \delta \left ( Y_1 + Y_2 - 2X - t\right ) \cr
& \cdot \exp \left \{ e^{-Y_1} + e^{-Y_2} + e^{-X}\right \}.
\end{align}
Next, we can integrate the field $X$. Up to factors to be removed by the normalisation, the delta yields to the following result :
\begin{align}
\mathcal{W}_{\mathbb{WCP} ^{1|1}_{(1,1 | 2)}} = \int (\mathcal{D} Y_1 \mathcal{D}
Y_2 ) e^{-Y_1 /2 - Y_2/2} \exp \left \{ e^{-Y_1} + e^{-Y_2} + e^{-Y_1/2 -Y_2/2 + t/2} \right \}.
\end{align}
We can then define the new variables
\begin{align}
y_i = e^{-Y_i /2}, \qquad i = 1, 2.
\end{align}
The measure changes as $- \frac{1}{2} y^{-1}_i \mathcal{D} y_i =
\mathcal{D} Y_i$, and therefore, up to factors in the normalisation, one gets
\begin{align}
\mathcal{W}_{\mathbb{WCP} ^{1|1}_{(1,1 | 2)}} = \int (\mathcal{D} y_1 \mathcal{D}
y_2 ) \exp \left \{ y_1^2 + y_2^2 + e^{t/2} y_1 y_2 \right \}.
\end{align}
One can then state that in the case of $\mathbb{WP} _{(2)}^{1|1}$ one does not get directly a geometry.  However, we can further introduce the new variables $\lambda$ and $x$, defined by
\begin{align}
y_1=y_2 x,\qquad\quad y_2^2=\lambda,
\end{align}
in such a way that, omitting an inessential constant factor, the final result can be achieved :
\begin{align}
\mathcal{W}_{\mathbb{WCP} ^{1|1}_{(1,1 | 2)}} = \int (\mathcal{D} x \mathcal{D} \lambda ) \exp \left \{\lambda \left( x^2 + 1 + e^{t/2} x\right) \right \}.
\end{align}
Thus, $\lambda$ is a multiplier and the geometric phase reduces to two points parametrized by $t$. This is a zero dimensional bosonic model in accordance with the results of Schwarz \cite{Schwarz-1995}.


\section{Conclusions}

\label{sec: conclusions} In the present paper we have investigated some basic questions about super Calabi-Yau varieties (SCY's). We have introduced a very general definition of a SCY, which encompasses a large class of
varieties, including the usual Calabi-Yau manifolds and several projective super spaces. We then restricted our analysis to the SCY with complex bosonic dimension $1$, proving that - beyond the usual elliptic curves - it contains the class of $N=2$ Super Riemann Surfaces (SRS's) and the projective super spaces $\proj{1|2}$ and $\mathbb{WP}_{(2)}^{1|1}$. As a byproduct of the mirror map construction, we realised at the very end that $\proj{1|2}$ is indeed a $N=2$ SRS: this provides a concrete realisation of a $N=2$ SRS by a map - the mirror map - into the Cartesian product of two copies of $\proj{1|1}.$ A comment is in order here. In the present paper we have referred to \cite{Witten-1} for the definition of $N=2$ SRS: in this case, the proof of triviality of the Berezinian bundle is given in \cite{Bei-1}. Nevertheless, there exists a more general definition of $N=2$ SRS given in \cite{FaRe-1990}. To the best of our knowledge, it is not completely obvious that the two definitions do actually coincide: indeed the definition of $N=2$ SRS in \cite{FaRe-1990} includes the definition in \cite{Witten-1} and, as a consequence, this should imply that all the $N=2$ SRS's in \cite{Witten-1}, \cite{Bei-1} and \cite{Bei-2} are holomorphically \emph{split}. Still, we feel like this topic deserve some more study.\\
Next, we have computed the super cohomology groups, which include integral forms, showing that for extended supersymmetric varieties a puzzle arises: when the picture number is not maximal nor vanishing, then the corresponding \v{C}ech cohomology groups are infinitely generated. Surely, this result will deserve a much deeper investigation; for instance, it would be interesting to understand if it enjoys a geometrical interpretation. Anyway, remarkably, we have shown that this kind of pathology is cured whenever one considers the de Rham cohomology of superforms, which is always finite, even when the corresponding group in \v{C}ech cohomology is infinite-dimensional. The same phenomenon occurs in arbitrary dimension $n|m$ as we have seen by explicitly
computing the de Rham cohomology of $\mathbb P^{n|m}$.
The computation of the sheaf cohomology also allowed us to determine the automorphisms of $\mathbb{P}^{1|2}$ and $\mathbb{WP}_{(2)}^{1|1}$, which, on the other hand, are rigid manifolds. It is interesting to note that for SCY with fermionic dimension larger than 1, the automorphism supergroup is never larger than the superprojective group. As announced, a more systematic analysis of the automorphism group will be presented in a separate paper. Finally, we have applied the mirror map defined by Aganagic and Vafa in \cite{AgaVafa}, showing that $\mathbb{P}^{1|2}$ is self-mirror (and, indeed, mapped to itself), whereas $\mathbb{WP}_{(2)}^{1|1}$ is mapped to a zero dimensional bosonic model. \newline
Even though we have chosen to investigate an apparently elementary framework, we realize that highly non-trivial aspects appear and some questions remains unanswered. For example, we have not been able to provide a suitable definition of K\"{a}hler structure (or K\"{a}hler moduli space) for SCY varieties. On one hand, SCY's of bosonic dimension $n=1$ having $\proj {1}$ as reduced space are simple enough in order to allow a complete analysis, as well as to shed some light on new interesting
properties of supermanifolds; on the other hand, they are too simple for providing a rich list of examples hinting to suitable solutions to the unanswered questions. The natural prosecution would then be to include properly the whole class of $N=2$ super Riemann surfaces, that are indeed SCY's with bosonic dimension $1$, and, more interestingly, to analyse SCY's with bosonic dimension $2$, i.e. \emph{super K3 varieties}. \\
Despite the results discussed above, we still cannot take our definition of SCY manifold as a definitive one. At the moment, indeed, the triviality of the Berezinian bundle alone appears as a provisional condition, maybe allowing for too many varieties to belong to the class. From this point of view, our definition might be considered as a pre-SCY condition. In this context, one might wonder whether the existence of a Ricci-flat metric is a natural condition to add, but in some meaningful example, such as $\mathbb{WP}_{(2)}^{1|1}$, it does not even exist. This seemingly suggests that Ricci-flatness is not the natural condition to add to the triviality of the Berezinian bundle.
These and other topics are currently under investigation.


\section*{Acknowledgments}
SN would like to thank Ron Donagi for having suggested this stimulating research topic.
SN and SLC would like to thank Gilberto Bini and Bert van Geemen for valuable discussions. \\
AM and RR would like to thank the Department of Science and High Technology, Universit\`{a} dell'Insubria at Como, and the Departments of Mathematics and
Physics, Universit\`{a} di Milano, for kind hospitality and inspiring environment.

\newpage \appendix

\section{Super Fubini-Study Metric and Ricci Flatness of $\pmb{\mathbb{P}^{1|2} }$}

\noindent We take on the computation of the super Ricci tensor for $\mathbb{P}^{1|2} $ starting from the local form, say in $U_z$, of the K\"ahler potential, given by
\begin{eqnarray}
K^{s} = \log (1 + z \bar z + \theta_1 \bar \theta_1 + \theta_2 \bar \theta_2).
\end{eqnarray}
This can of course be expanded in power of the anticommuting variables as in \cite{Sethi}, but it is not strictly necessary to our end. \newline
In the following we will adopt this convention: we use latin letters $i, j, \ldots $ for bosonic indices, Greek letters $\alpha, \beta, \ldots $ for fermionic indices and capital Latin letters $A, B, \ldots $ will gather both
of them. The convention on the unbarred and barred indices goes as usual.\newline
The \emph{holomorphic} and \emph{anti-holomorphic} super derivatives are defined as follows (in the local patch):
\begin{eqnarray}
\partial \defeq \partial_z dz + \partial_{\theta_\alpha} d\theta_\alpha, \qquad \quad \bar \partial \defeq \partial_{\bar z} d\bar z + \partial_{\bar
\theta_{\bar \alpha}} d\bar \theta_{ \bar \alpha},
\end{eqnarray}
where $\alpha, \bar \alpha = 1,2$: in other words we have $\partial \defeq \partial_A d X^A$ and $\bar \partial \defeq \partial_{\bar A} d \bar X^{\bar A}$ with $dX^A = (dz | d\theta_1, d\theta_2)$ and $d\bar X^{\bar A} = (d \bar z | d \bar \theta_1, d \bar \theta_2)$.
It is important to stress that while the holomorphic derivative $\partial $ acts as usual from the left to the right, the anti-holomorphic derivative $\bar \partial$ acts from the right to the left instead (even if it is written on left of the function
acted on). We also stress that $\partial$ and $\bar \partial$ behave as a standard exterior derivative $d$ on forms. As such the derivatives ``do not talk'' at all with the forms and only acts on functions, while the forms in $\partial $ or $\bar \partial$ are
moved to the right and in turn do not talk to the functions acted by the derivatives. This means that, for example, considering the local expression for a (holomorphic) $1$-form acted on by $\partial$, we will find:
\begin{eqnarray}
\partial (f(z|\theta_1, \theta_1) d\theta_1) = (\partial_{B} f(z|\theta_1, \theta_1)) dX^B d\theta_1.
\end{eqnarray}
Coherently, we will \emph{never} consider expression of the kind $dX^B f(z|\theta) d\theta_1$, so that we will never have to commute or anti-commute a form with a function to get $\pm f(z|\theta) dX^B d\theta_1$:
forms and functions just don't talk to each other and the form in $\partial $ and $\bar \partial$ are moved the right.\newline We now define the super K\"ahler form as
\begin{eqnarray}
\Omega^s \defeq \partial \bar \partial K^s \quad \mbox{or analogously} \quad \Omega^s = \partial_A \partial_{\bar B } K^{s} dX^A d\bar X^{\bar B}.
\end{eqnarray}
The super metric tensor $H^s_{A\bar B}$ can then be red out of it, similarly to the ordinary complex geometric case:
\begin{eqnarray}
H^s_{A \bar B} = \partial_A \partial_{\bar B} K^s.
\end{eqnarray}
We now deal with the derivative of the super K\"ahler potential $K^s$. Remembering that $\partial_{\bar B}$ acts from the right, it is straightforward to check that:
\begin{eqnarray}
\partial_{\bar B} K^s = \partial_{\bar B} \log (1 + z \bar z + \theta_1 \bar \theta_1 + \theta_2 \bar \theta_2) = \frac{z d\bar z + \theta_1 d \bar
\theta_1 + \theta_2 d \bar \theta_2 }{1+ z \bar z + \theta_1 \bar \theta_1 + \theta_2 \bar \theta_2}.
\end{eqnarray}
We now have a product of functions: since we are dealing with anti-commuting objects we need to make a careful use of the generalized Leibniz rule
\begin{eqnarray}
\partial (f \cdot g ) = ( \partial f ) \cdot g + (-1)^{|\partial| |f|} f \cdot (\partial g).
\end{eqnarray}
We will put $f \defeq z d\bar z + \theta_1 d \bar \theta_1 + \theta_2 d \bar \theta_2 $ and $g \defeq 1/(1+ z \bar z + \theta_1 \bar \theta_1 + \theta_2 \bar \theta_2)$ in the following computation. \newline
While the first bit of the $\partial$ derivative is pretty straightforward and simply gives
\begin{eqnarray}
( \partial f ) \cdot g = \frac{1}{1+ z \bar z + \theta_1 \bar \theta_1 + \theta_2 \bar \theta_2} \left ( dz d \bar z + d\theta_1 d\bar \theta_1 + d\theta_2 d\bar \theta_2 \right ),
\end{eqnarray}
the second contribution need some extra care: to avoid errors, we may split the derivatives in $\partial$ by linearity, bearing in mind the non-trivial commutation relation in the generalised Leibniz rule above. \newline
We have the following contribution from $(-1)^{|\partial| |f|} f \cdot (\partial g)$:
\begin{align}
& \partial_z \left ( \frac{z}{1+ |z|^2 + \theta^2 } \right ) dz d\bar z +\partial_z \left ( \frac{\theta_1}{1+ |z|^2 + \theta^2 } \right ) dz d\bar \theta_1 + \partial_z \left ( \frac{\theta_2}{1+ |z|^2 + \theta^2 } \right ) dz d\bar \theta_2  \notag
\end{align}
\begin{eqnarray}
= \frac{-|z|^2 dz d\bar z - \theta_1 \bar z dz d\bar \theta_1 - \theta_2 \bar z dz d\bar \theta_2}{\left ( 1+ |z|^2 + \theta^2 \right )^2}
\end{eqnarray}
where we have written $\theta^2 \defeq \theta_1 \bar \theta_1 + \theta_2 \bar \theta_2$. Notice, incidentally that the minus signs above do not come from the commutation relation, but just from the derivative: the commutation
relation gives contribution when $\partial_{\theta_i}$ is involved
\begin{align}
& \partial_{\theta_1} \left ( \frac{z}{1+ |z|^2 + \theta^2 } \right ) d\theta_1 d\bar z + \partial_{\theta_1} \left ( \frac{\theta_1}{1+ |z|^2 +
\theta^2 } \right ) d\theta_1 d\bar \theta_1 + \partial_{\theta_1} \left ( \frac{\theta_2}{1+ |z|^2 + \theta^2 } \right ) d\theta_1 d\bar \theta_2 \notag
\end{align}
\begin{eqnarray}
= \frac{-z \theta_1 d\theta_1 d\bar z + \theta_1 \bar \theta_1 d\theta_1 d\bar \theta_1 + \theta_2 \bar \theta_1 d\theta_1 d\bar \theta_2}{\left ( 1+|z|^2 + \theta^2 \right )^2},
\end{eqnarray}
\begin{align}
& \partial_{\theta_2} \left ( \frac{z}{1+ |z|^2 + \theta^2 } \right ) d\theta_2 d\bar z + \partial_{\theta_2} \left ( \frac{\theta_1}{1+ |z|^2 + \theta^2 } \right ) d\theta_2 d\bar \theta_1 + \partial_{\theta_2} \left (
\frac{\theta_2}{1+ |z|^2 + \theta^2 } \right ) d\theta_2 d\bar \theta_2 \notag
\end{align}
\begin{eqnarray}
= \frac{-z \theta_2 d\theta_2 d\bar z + \theta_2 \bar \theta_1 d\theta_2 d\bar \theta_1 + \theta_2 \bar \theta_2 d\theta_2 d\bar \theta_2}{\left ( 1+ |z|^2 + \theta^2 \right )^2}.
\end{eqnarray}
Putting together all the pieces we have:
\begin{align}
\partial \bar \partial K^s = \frac{1}{\left ( 1+ |z|^2 + \theta^2 \right )^2} & \Big [ (1+\theta^2) dz d \bar z + (1+ |z|^2 + 2 \theta_1 \bar \theta_1 + \theta_2 \bar \theta_2) d\theta_1 d \bar \theta_1 +  \notag \\
& + (1+ |z|^2 + \theta_1 \bar \theta_1 + 2 \theta_2 \bar \theta_2) d\theta_2 d \bar \theta_2 - \theta_1 \bar z dz d \theta_1 +  \notag \\
& - \theta_2 \bar z dz d \theta_2 - z \bar \theta_1 d\theta_1 d\bar z - z \bar \theta_2 d\theta_2 d\bar z + \theta_2 \bar \theta_1 d\theta_1 d\bar
\theta_2 + \theta_1 \theta_2 d\theta_2 d\bar \theta_1 \Big ],  \notag
\end{align}
so the supermetric reads
\begin{eqnarray}
H^s_{A\bar B} = \left ( \begin{array}{c|cc}
1 + \theta^2 & - \theta_1 \bar z & - \theta_2 \bar z \\ \hline
&  &  \\
- z \bar \theta_1 & 1+ |z|^2 + 2 \theta_1 \bar \theta_1 + \theta_2 \bar \theta_2 & \theta_2 \bar \theta_1 \\
&  &  \\
- z \bar \theta_2 & \theta_1 \bar \theta_2 & 1+ |z|^2 + \theta_1 \bar \theta_1 + 2 \theta_2 \bar \theta_2%
\end{array}
\right ).
\end{eqnarray}
Using the metric one can generalise the expression for the Ricci tensor one has in ordinary complex geometry, by substituting the determinant with the Berezinian:
\begin{eqnarray}
\mbox{Ric}_{A\bar B} = \partial_A \partial_{\bar B} \log \left ( \mbox{Ber} \, H^s \right ).
\end{eqnarray}
So the first thing we need to evaluate to prove the (super) Ricci flatness of $\mathbb{P}^{1|2} $ is the Berezinian of the super metric above. \newline
We recall that in general, considering a generic square matrix $X$ valued in a super commutative ring, we have
\begin{eqnarray}
\mbox{Ber}(X) = \det(A) \det (D- C A^{-1}B)
\end{eqnarray}
where $A, B, C, D$ are the blocks as enlightened above. Notice that $A$ and $D$ are \emph{even} while $B$ and $C$ are \emph{odd}. \newline
We underline that in our case, to make sense out of the expression above we have to look at $C A^{-1}B$ as Kronecker product, as follows:
\begin{eqnarray}
CA^{-1}B \rightarrow A^{-1} \cdot C \otimes B,
\end{eqnarray}
$A^{-1}$ consisting of a single even element. \newline
We start from the computation of $A^{-1} \cdot C \otimes B$. We have:
\begin{eqnarray}
A^{-1} = \left (\frac{1+ \theta^2}{(1+ |z|^2 + \theta^2)^2}\right)^{-1},
\end{eqnarray}
\begin{eqnarray}
C= \frac{1}{(1+ |z|^2 + \theta^2)^2}\left (
\begin{array}{c}
- z \bar \theta_1 \\
- z \bar \theta_2
\end{array}
\right ),
\end{eqnarray}
\begin{eqnarray}
B = \frac{1}{(1+ |z|^2 + \theta^2)^2} \left ( - \theta_1 \bar z , \; - \theta_2 \bar z \right ).
\end{eqnarray}
This leads to:
\begin{align}
A^{-1}\cdot C \otimes B & = \frac{1}{(1+ \theta^2)(1+ |z|^2 + \theta^2)^2} \left (
\begin{array}{c}
- z \bar \theta_1 \\
- z \bar \theta_2
\end{array}
\right ) \otimes \left ( - \theta_1 \bar z , \; - \theta_2 \bar z \right )  \cr & = - \frac{|z|^2}{(1+ \theta^2)(1+ |z|^2 + \theta^2)^2} \left (
\begin{array}{cccc}
\theta_1\bar \theta_1 &  &  & \theta_2 \bar \theta_1 \\
&  &  &  \\
\theta_1 \bar \theta_2 &  &  & \theta_2 \bar \theta_2
\end{array}
\right )
\end{align}
where the overall minus sign comes from the commutation relation of the theta's. It is actually convenient to multiply $(1+\theta^2)^{-1}$ out: first of all we observe
\begin{eqnarray}
\frac{1}{(1+\theta^2)} = 1-\theta^2 + 2\theta^4
\end{eqnarray}
where $\theta^4 \defeq \theta_1 \bar \theta_1 \theta_2 \bar \theta_2$. So the product above becomes:
\begin{eqnarray}
A^{-1}\cdot C \otimes B = -\frac{|z|^2}{(1+ |z|^2 + \theta^2)^2} \left (
\begin{array}{cccc}
\theta_1\bar \theta_1 - \theta^4 &  &  & \theta_2 \bar \theta_1 \\
&  &  &  \\
\theta_1 \bar \theta_2 &  &  & \theta_2 \bar \theta_2 -\theta^4
\end{array}
\right ).
\end{eqnarray}
Therefore one has the following expression:
\begin{align}
&D- CA^{-1}B = \frac{1}{(1+ |z|^2 + \theta^2)^2}  \nonumber \\
&\cdot \left ( \begin{array}{cccc}
1+ |z|^2 + (2+ |z|^2) \theta_1 \bar \theta_1 + \theta_2 \bar \theta_2 - |z|^2 \theta^4  & (1+|z|^2) \theta_2 \bar \theta_1 \\ \\
(1+|z|^2) \theta_1 \bar \theta_2 & 1+ |z|^2 + \theta_1 \bar \theta_1 + (2+ |z|^2) \theta_2 \bar \theta_2 - |z|^2 \theta^4
\end{array}\right ).   \nonumber
\end{align}
We now need to evaluate the determinant of the square matrix above:
\begin{align}
\det (D-&CA^{-1}B) = \frac{1}{(1+ |z|^2 + \theta^2)^4} \Big [ (1+ |z|^2)^2 +(1+|z|^2)\theta_1 \bar \theta_1\notag \\
&  + (1+|z|^2)(2+|z|^2) \theta_2\bar\theta_2 + (1+|z|^2)(2+|z|^2)\theta_1 \bar \theta_1 + (1+ |z|^2) \theta_2 \bar \theta_2 +  \notag \\
& + \theta^4 + (2 + |z|^2)^2 \theta^4 - 2 |z|^2(1+ |z|^2) \theta^4 + (1 + |z|^2)^2 \theta^4 \Big ]
\end{align}
where we have isolated on different lines the zeroth, quadratic and quartic contribution in the theta's. We can simplify a little the expression above to get:
\begin{align}
\det (D-CA^{-1}B) & =\frac{(1+|z|^2)^2}{(1+ |z|^2 + \theta^2)^4} \Big [ 1 +\frac{3+|z|^2}{1+|z|^2} \theta^2 + \frac{6 + 4|z|^2}{(1+|z|^2)^2} \theta^4 \Big ].
\end{align}
To evaluate the full Berezinian we need to invert the determinant we just got. This yields:
\begin{align}
\frac{1}{\det (D-CA^{-1}B)} & = (1+ |z|^2 + \theta^2)^4 \Big [ 1 - \frac{3+|z|^2}{1+|z|^2} \theta^2 - 2 \frac{6 + 4 |z|^2 + |z|^4}{(1+ |z|^2)^2}\theta^4 \Big ].
\end{align}
Putting together the pieces, we can evaluate the full Berezinian:
\begin{align}
\mbox{Ber} (H^s) & = \frac{(1+ |z|^2 + \theta^2)^2 (1+ \theta^2 )}{(1+|z|^2)^2} \Big [ 1 - \theta^2 - \frac{2}{1+ |z|^2} \theta^2 - 2 \frac{6 + 4 |z|^2 + |z|^4}{(1+ |z|^2)^2}\theta^4 \Big ] = 1.  \notag
\end{align}
Remembering that $\mbox{Ric}_{A\bar B} = \partial_A \partial_{\bar B} \log (\mbox{Ber} (H^s))$, since we have found that $\mbox{Ber} (H^s) =1$, this
leads us the the conclusion:
\begin{eqnarray}
\mbox{Ric}_{A\bar B} = 0.
\end{eqnarray}
$\mathbb{P}^{1|2} $ is Ricci-flat and therefore it is a super Calabi-Yau manifold in the strong sense.


\end{document}